\documentclass[pre,aps,floatfix,a4paper,reprint]{revtex4-1}

\usepackage{ulem}
\usepackage{graphicx}
\usepackage{amsmath}
\usepackage{float}
\usepackage{placeins}
\usepackage{bm}
\usepackage{xcolor}

\begin{document}

\title{
 Mechanisms for recirculation cells in granular flows in rotating cylindrical rough tumblers}

\author{Umberto D'Ortona}
\email{umberto.d-ortona@univ-amu.fr}
\affiliation{Aix Marseille Univ., CNRS, Centrale Marseille, M2P2, Marseille, France}
\author{Nathalie Thomas} 
\affiliation{Aix Marseille Univ., CNRS, IUSTI, Marseille, France}
\author{Richard M. Lueptow}
\affiliation{Department of Mechanical Engineering, 
         Northwestern University, Evanston, Illinois 60208, USA}
\affiliation{Department of Chemical and Biological Engineering, Northwestern University, Evanston, Illinois 60208, USA}

\date{\today}
\begin{abstract} 
    Friction at the endwalls of partially-filled horizontal
	rotating tumblers induces curvature and axial drift of particle
	trajectories in the surface flowing layer. Here we describe the results
	of a detailed discrete element method study of the dry granular flow of
	monodisperse particles in three-dimensional cylindrical tumblers with
	endwalls and cylindrical wall that can be either smooth or rough. 
	Endwall roughness induces more curved particle trajectories, while a smooth cylindrical wall enhances drift near the endwall.  This drift induces recirculation cells near the endwall.  The use of mixed roughness (cylindrical wall and
	endwalls having different roughness) shows the influence of each wall on the drift and curvature of particle trajectories as well as the modification of the free surface topography.  The effects act in
	opposite directions and have variable magnitude along the length of the tumbler such that their sum determines both direction of net drift and the recirculation
	cells. Near the
	endwalls, the dominant effect is always the endwall effect, and the axial drift for surface particles is toward the
	endwalls.  For long enough tumblers, a counter-rotating cell occurs adjacent to each of the endwall cells having a surface drift toward the center because the cylindrical wall effect is dominant there. These cells are not
	dynamically coupled with the two endwall cells.  The competition between the drifts induced by
	the endwalls and the cylindrical wall
	determines the width and drift amplitude for both types of cells.
\end{abstract}
\maketitle

\section{Introduction}

Horizontal cylindrical rotating tumblers that are partially filled with
granular materials are considered a canonical
system for the study of flowing granular materials.
They are commonly used in industrial processes for mixing,
coating, and granulation.
Although granular
flow in cylindrical tumblers has long been
studied \cite{GDRMidi04,OttinoKhakhar00,MeierLueptow07},
much research
has focused on quasi-two-dimensional
circular tumblers \cite{OrpeKhakhar04,JainOttino02,OrpeKhakhar07,ClementRajchenbach95,FelixFalk02,FelixFalk07}, 
where friction at the endwalls 
plays an important role \cite{PohlmanOttino06}. 
Even in longer tumblers, the flow near the flat endwalls of the
tumbler is strongly affected by the friction between the endwalls and
the flowing particles \cite{PohlmanOttino06,SantomasoOlivi04,ChenOttino08,DOrtonaThomas18}.
Moreover, the endwall
boundary affects the mixing of monodisperse particles \cite{SantomasoOlivi04} and plays a
role in initializing axial segregation bands of bidisperse
particles \cite{BridgwaterSharpe69,HillKakalios94,FiedorOttino03} near the endwalls.
In this paper, we consider the role of wall roughness of both
the cylindrical wall and the endwalls for granular flow in 
cylindrical tumblers.

In long cylindrical tumblers with geometrically smooth but frictional walls,
not only is the streamwise velocity near the tumbler
endwall slower than that at the center of the
tumbler\ \cite{ManevalHill05}, but the
resulting reduced mass transport induces a local axial flow near the endwalls
\cite{SantomasoOlivi04,ChenOttino08}. 
Particles near the endwall flow down the slope more slowly
than particles far from the endwall. As a consequence, to conserve mass
they flow axially away from the
endwall in the upper portion of the flowing layer and back toward the
endwall in the downstream portion of the flowing layer
\cite{ManevalHill05,PohlmanOttino06,PohlmanMeier06,DOrtonaThomas18}.  
For a half-filled tumbler, the region that
is affected by endwall friction (where particle trajectories are curved) 
extends about a radius of the tumbler 
from the endwall
\cite{PohlmanMeier06,ChenOttino08}. 
A second effect of the endwall friction is the existence of a pair of
recirculation 
cells next to the tumbler endwalls, and, if the tumbler is long 
enough, a second pair of counter-rotative cells in between the endwall cells
and the midlength of the tumbler \cite{DOrtonaThomas18}. These cells are called ``central cells," even though for very long tumblers they  do not extend to the center of the tumbler but remain adjacent to the endwall cells.

Analogous recirculation cells also appear in  spherical and double-cone 
tumblers, except that only one recirculation cell appears on either 
side of the equator \cite{ZamanDOrtona13,DOrtonaThomas15}.
Recent studies of granular flow in a spherical tumbler indicate that the wall
roughness, either a geometrically smooth wall or a wall made up of particles,
can strongly influence the recirculation cells for monodisperse particles
flows \cite{DOrtonaThomas15}
and the segregation pattern for bidisperse particles flows \cite{ChenLueptow10,DOrtonaThomas16}.
Regardless of wall roughness, particles drift axially toward
the pole near the surface of the flowing layer with a return flow toward
the equator that occurs deeper in the flowing layer resulting in a
global circulation of granular material.  The recirculation
cells are quite difficult to observe.
The axial drift induced by the recirculation cells 
in a 0.14~m diameter tumbler with 2~mm flowing particles
is typically only about 1-2 millimeters each time a particle passes through
the flowing layer.  Consequently, the axial drift and resulting
recirculation cells are buried in the noise related 
collisional diffusion as particles flow. 
Nevertheless, axial drift affects segregation 
pattern formation in spherical tumblers with size-bidisperse particles \cite{DOrtonaThomas16,YuLueptow20}.  

Both experiments and DEM
simulations indicate that the recirculation cells in spherical and conical
tumblers result from the combination of wall friction and tumbler geometry. Specifically, 
the roughness of the tumbler wall, varied by comparing a geometrically smooth 
wall with a wall constructed of particles, 
affects the degree of axial drift and
the thickness of the flowing layer near the walls
\cite{DOrtonaThomas15}. 
The shape of the wall (spherical versus conical) also alters the
axial drift 
as a result of the variation in the length of the flowing layer \cite{ZamanDOrtona13} and the resulting variation in flux injected into the flowing layer from the bed of particles in solid body rotation.

In other flow geometries, secondary flows organized as recirculation cells
also occur. In a horizontal channel with moving lateral walls, 
two longitudinal counter-rotating recirculation cells form 
\cite{KrishnarajNott15}. Due to these cells, the particles rise at
the center of the channel and move down next to the wall. In a granular cylindrical
Couette cell, only one recirculation cell appears, and it is localized next to the
inner cylinder. 
The particles move down along the inner cylindrical wall for both smooth and
rough walls \cite{KrishnarajNott15}. 
In an inclined channel with smooth walls, two or four recirculation
cells are obtained depending on the thickness of the flow and 
the tilt angle \cite{BroduRichard13,BroduDelannay15}.
For wide channels and steeper inclines (more
than 30$^\circ$), recirculation
cells appear without any lateral wall \cite{ForterrePouliquen01,ForterrePouliquen02,BorzsonyiEcke09}. Finally, for a bidisperse granular flow of particles with different sizes and densities
down an incline, segregation and the Rayleigh-Taylor instability combine 
to induce recirculation cells analogous to Rayleigh-B\'enard convection cells
\cite{DOrtonaThomas20}.

In this paper, we study recirculation cells in a cylindrical tumbler with 
different degrees of wall roughness using the
discrete element method (DEM) \cite{CundallStrack79,Ristow00,SchaferDippel96}.
In the cylindrical tumbler geometry, the flux of particles entering and leaving the flowing layer is nearly constant along the entire length of 
the tumbler, so the impact of its variation on the recirculation cell that occurs in spherical and 
double-cone tumblers is eliminated. In this way, the 
effect of wall roughness alone can be clarified.
Our goal is to examine the effect of wall roughness 
on the axial drift 
and the resulting recirculation cells, which is likely crucial
to understand the mechanisms of mixing of monodisperse particles
\cite{SantomasoOlivi04}
and the initiation of axial bands of segregated bidisperse particles
\cite{BridgwaterSharpe69,DonaldRoseman62,DasguptaKhakhar91,Nakagawa94,HillKakalios94,HillKakalios95,HillCaprihan97,FiedorOttino03} in the endwall regions. 

\section{DEM Simulations}

For the
DEM simulations, a standard linear-spring and viscous damper force model
\cite{ChenOttino08,SchaferDippel96,Ristow00,CundallStrack79} is used to calculate
the normal force between two contacting particles: 
${\bm F}_n^{ij}=[k_n\delta - 2 \gamma_n m_{\rm eff} ({\bm V}_{ij} \cdot {\bm{\hat r}_{ij}})]{\bm{\hat r}_{ij}}$, 
where $\delta$ and $\bm V_{ij}$ are the particle overlap and the relative
velocity $(\bm V_i - \bm V_j)$ of contacting particles $i$ and $j$ respectively; 
$\bm{\hat r}_{ij}$ is
the unit vector in the direction between particles $i$ and $j$; 
$m_{\rm eff} = m_i m_j/(m_i + m_j)$ is the reduced mass of the two particles; 
$k_n = m_{\rm eff} [( \pi/\Delta t )^2 + \gamma^2_n]$ is the normal stiffness 
and $\gamma_n = \ln e/\Delta t$ is the normal damping, where $\Delta t$
is the collision time and $e$ is the restitution coefficient \cite{ChenOttino08,Ristow00}. A
standard tangential force model \cite{SchaferDippel96,CundallStrack79} with
elasticity is implemented: $\bm F^t_{ij}= -\min(|\mu F^n_{ij}|,|k_s\zeta|){\rm
sgn}(V^s_{ij})\,\bm{\hat s}$, where
$V^s_{ij}$ is the relative tangential velocity of two particles \cite{Rapaport02},
$k_s$ is the tangential stiffness, $\mu$ the Coulomb friction coefficient, $\zeta(t) = \int^t_{t_0} V^s_{ij} (t') dt'$ is the net
tangential displacement after contact is first established at time $t =
t_0$, and ${\bm{\hat s}}$ is the unit vector in the tangential direction.
The velocity-Verlet algorithm \cite{Ristow00,AllenTildesley02} is used to 
update
the position, orientation, and linear and angular velocity of each
particle. Tumbler walls (cylindrical wall and endwalls) are modeled either as smooth frictional surfaces (smooth wall) or as a monolayer of bonded particles of
different diameters to vary the roughness (rough walls). Both wall conditions
have infinite mass for calculation of the collision force between the
tumbling particles and the wall.

The horizontal cylindrical tumblers considered here have a radius $R=0.07$~m and lengths varying
from $L=0.07$ to 0.42~m{\color{black}, corresponding to $R/d=35$ and $L/d=35$ to 210, where $d$ is the particle diameter, and $L/2R=0.5$ to 3}. They are filled to volume fractions (fill levels) from 20\% to 
50\% with $d=2$~mm particles with particle
properties that correspond to cellulose
acetate: density $\rho =$ 1308~kg~m$^{-3}$, restitution coefficient $e = 0.87$ \cite{DrakeShreve86,FoersterLouge94,SchaferDippel96}. The particles are initially randomly
distributed in the tumbler with a total number of particles 
ranging from about $2 \times 10^4$ to $2.7 \times 10^6$.
To avoid a close-packed structure, the particles have a
uniform size distribution ranging from 0.95$d$ to 1.05$d$. 
The friction
coefficient between particles and between particles and walls is set to 
$\mu = 0.7$. 
Gravitational acceleration is $g$ = 9.81~m~s$^{-2}$, 
the collision time is $\Delta t$ =10$^{-4}$ s, consistent with previous
simulations \cite{TaberletNewey06,ChenLueptow11,ZamanDOrtona13} and sufficient for
modeling hard spheres \cite{Ristow00,Campbell02,SilbertGrest07}.  These
parameters correspond to a stiffness coefficient $k_n = 7.32\times 10^4$ (N m$^{-1}$)
\cite{SchaferDippel96} and a damping coefficient $\gamma_n = 0.206 $~kg~s$^{-1}$.  The
integration time step is $\Delta t/50 = 2\times 10^{-6}$~s to meet the requirement of
numerical stability \cite{Ristow00}.

\begin{figure}[htbp]
\includegraphics[width=0.99\linewidth]{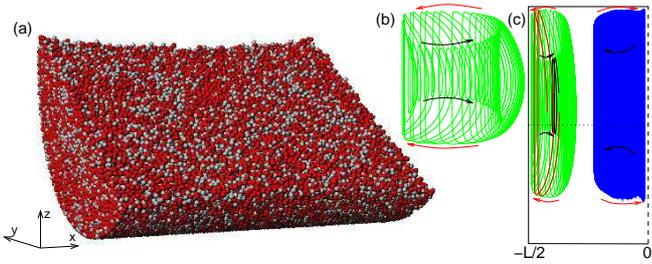}
\caption{(a) Granular flow in a 0.14~m long {\color{black}($L/d=70$, $L/2R=1$)} and 0.14~m diameter cylindrical tumbler {\color{black}($R/d=35$)}
filled to 30\% by volume with {\color{black}$d=2$}~mm particles randomly colored red and gray. The
tumbler has smooth frictional walls, and the rotation speed is 15~rpm {\color{black}($Fr=\omega^2 R/g= 0.018$)}.
(b-c) Top view of the two recirculation cells obtained in the left half of the 
tumbler. The vertical dashed line indicates the center of the tumbler
and the horizontal dotted line denotes the axis of rotation. The endwall cell trajectory (green) 
is integrated over 200~s and corresponds to about 3 orbits through the cell,
 while for the central adjacent cell (blue), 500~s are required for only one orbit. {\color{black} In (b), the horizontal axis of the endwall recirculation cell is stretched compared to the vertical axis to clearly show the recirculation cells. In (c), two passes in
the flowing layer are highlighted in red for the outer trajectory and in black for the inner
trajectory. The corresponding arrows in (b, c) show the
recirculation direction. (Only the arrows are shown for the
central recirculation cell.)}}
\label{figintro}
\end{figure}
{\color{black}The tumbler typically rotates at 15 rpm, corresponding to a Froude number $Fr=\omega^2 R/g= 0.018$, although a range of rotation speeds, 2.5 to 30 rpm ($0.0005\le Fr\le 0.070$), are also considered.  This range of $Fr$ corresponds to the continous-flow rolling regime for tumbler flow, characterized by a steady flowing layer with a surface that is essentially flat~\cite{MeierLueptow07}.}  Figure \ref{figintro}(a) shows a typical simulation of monodisperse granular flow. Particles are randomly colored red and gray to improve the visualization. 
The $x$-axis is the axis of 
rotation  with the $z$-axis opposite to $\bm g$ and the $y$-axis perpendicular 
to $x$ and $z$. The origin is at the midlength of the tumbler, though it is
shown as offset in Fig.~\ref{figintro}(a).
The velocity throughout the entire domain is obtained
by binning particles in a 3D grid and averaging their velocity over 50~s 
of physical time (2.5 $\times$ 10$^7$ integration time steps) to assure 
an adequately smooth velocity field.
 The first 20~s of simulation are omitted to assure that 
the flow is steady before averaging.
Mean particle trajectories are obtained by integrating the particle 
velocity based
on the velocity field or by averaging particles trajectories based on 
particle positions stored every 0.1s. The two methods provide similar results {\color{black} and are consistent with measurements using an x-ray system to track the location of a single x-ray opaque tracer particle in an analogous experimental setup \cite{DOrtonaThomas15}}. 

Figure~\ref{figintro}(b) shows two trajectories integrated for 50 tumbler 
rotations
(green-left) or 125 rotations (blue-right) corresponding to the two 
left recirculation cells among the four cells of the tumbler \cite{DOrtonaThomas18}.
For the endwall trajectory (green),
the particle drift is toward the endwall for particles
near the flowing layer surface (outer spiral of the cell trajectory), and
it is toward the midlength of the tumbler for particles deeper in the
flowing layer  (inner spiral of the cell trajectory at the core of the
cell). The trajectory is opposite in the central cell (blue).
The recirculation cells have previously been described in detail \cite{DOrtonaThomas18} {\color{black} and confirmed experimentally using colored bands of particles and colored tracer particles in spherical tumblers \cite{ZamanDOrtona13}.} {\color{black} By considering individual particle trajectories, it is possible to compute the trajectory diffusion. For a typical system (0.14~m long tumbler filled to 30\% and rotating at 15~rpm), the standard deviation 
of the drift is 3.5~mm while
the maximal measured drift is around 2.5~mm and the maximal curvature is around 10~mm.}

\section{Effect of wall roughness}

\subsection{Particle trajectories and wall roughness}
\label{link2}
 
To study the axial drift of particles along the length of the 
tumbler, mean trajectories are integrated from the velocity field obtained
from the DEM simulations {\color{black} for cases where both the endwalls and cylindrical wall are smooth or both are rough}.
\begin{figure}[htbp]
\includegraphics[width=0.95\linewidth]{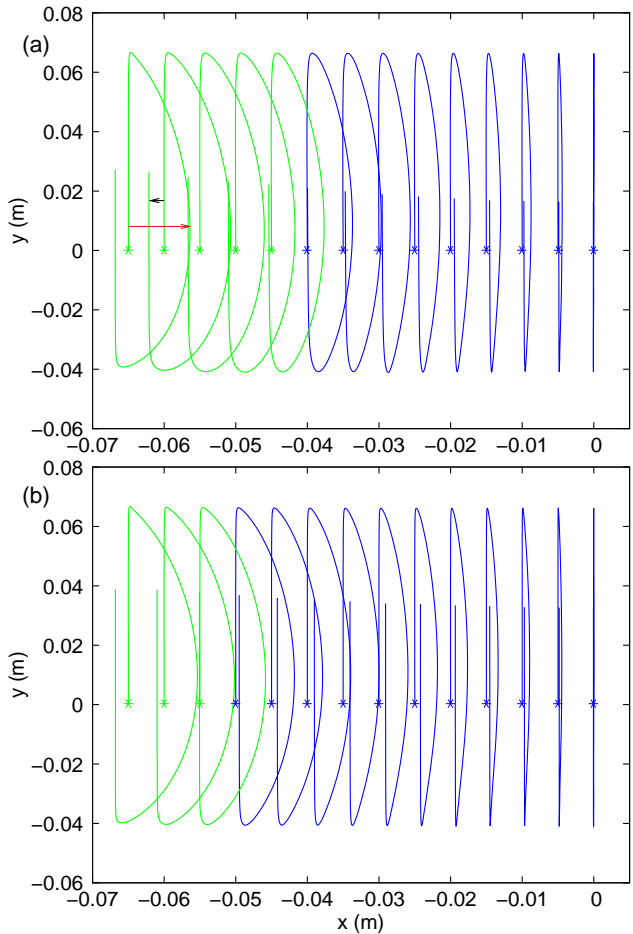}
\caption{Mean trajectories of 2~mm particles {\color{black} starting from the positions marked with a *, which are} equally spaced every 5~mm along a 
0.14~m {\color{black}($L/2R=1$) tumbler with} (a) smooth {\color{black} endwalls and cylindrical} wall or (b) 2~mm rough  {\color{black} endwalls and cylindrical} wall filled to 30\% and rotating at 
15~rpm {\color{black}($Fr= 0.018$)} (top view). 
Green trajectories have drift toward
the endwall and blue toward the center. 
Only the left half of the tumbler is shown. Note that the horizontal and vertical axes are scaled differently. All trajectories are integrated for 2.5~s 
to assure that they start and end in the static zone. {\color{black} The red arrow indicates the curvature, and the black
arrow indicates the drift.}} 
\label{cyl14traj}
\end{figure}
Figure~\ref{cyl14traj} shows average particle trajectories for one pass through
the flowing layer for different initial axial locations. The trajectories
start from the static zone where particles are in solid body rotation (marked with a star), 3~mm above the bottom cylindrical wall,
and in 
the $y=0$ plane that includes the axis of rotation. Particles starting in
this location will
follow a trajectory such that they are very near the surface when in the flowing layer. While in the
static zone, particle trajectories follow a straight vertical line 
until they reach
the flowing layer (uppermost point) and then flow down the slope following a curved trajectory, 
except at the center of the tumbler ($x=0$), which is a symmetry plane.
When they
reach the static zone (bottommost point), they again follow a straight vertical
line in the static zone. The {\sl curvature} of the trajectory is defined as the maximum axial 
displacement from the starting point (red arrow in Fig~\ref{cyl14traj}(a)).

 It is important to note that the start {\color{black} (marked by a *)} and end points for a single pass 
through the flowing layer do not coincide.  For particles near the endwall 
(green trajectories), the particles have a net axial displacement, or {\sl drift}, toward the endwall {\color{black}(black arrow in Fig.~\ref{cyl14traj}(a))},
 while particles further from the endwall of the tumbler (blue trajectories) 
drift toward the center (midlength) of the tumbler. The trajectories in 
Fig.~\ref{cyl14traj} correspond to the motion of particles near the surface 
of the flowing layer.  Particles deeper in the flowing layer have a 
 net drift in the opposite direction to conserve mass.  That is, particles deep in the 
flowing layer near the endwall drift toward the center of the tumbler, 
while particles near the center of the tumbler drift toward the endwall. 
In both cases, the axial drift of particles deep in the flowing layer 
balances the drift of particles near the surface resulting in the 
recirculation cells shown in Fig.~\ref{figintro}(b) \cite{DOrtonaThomas18}.

The general pattern of the curvature of the particle trajectories and the 
axial drift are similar, regardless of whether the tumbler walls (cylindrical wall
and endwall) are smooth 
(Fig.~\ref{cyl14traj}(a)) or both the cylindrical wall and endwalls are formed 
from a monolayer of 2~mm particles (Fig.~\ref{cyl14traj}(b)). In both cases 
the trajectories for particles near the surface of the flowing layer 
are curved with a maximum curvature near the endwalls, 
concave
to the endwalls. 
The main difference between rough and smooth
wall tumblers is the location of the transition from trajectories drifting 
toward the endwall (green) and
drifting toward the center (blue). The point where the drift changes sign indicates 
the boundary between the two recirculation
cells at the surface. It is clear that the endwall recirculation cells are smaller
for 2~mm rough walls than for a tumbler with smooth walls.

\begin{figure}[htbp]
\includegraphics[width=0.95\linewidth]{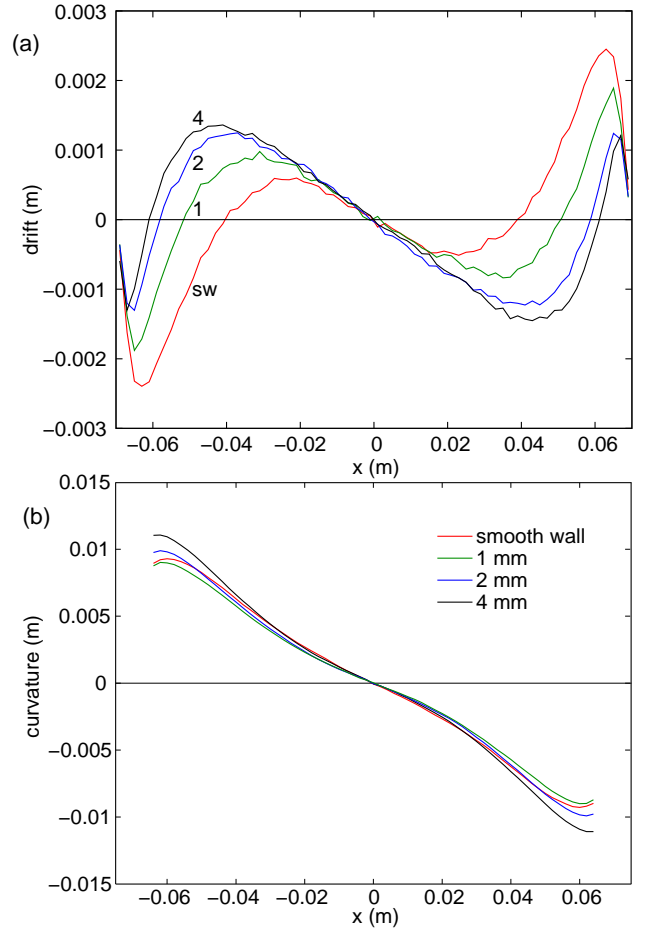}
\caption{(a) Axial drift near the surface and (b) trajectory curvature measured along the length of a 
30\% full 0.14~m {\color{black}($L/2R=1$)} long tumbler rotating at 15~rpm {\color{black}($Fr=0.018$)} for particles
starting 3~mm from 
the cylindrical wall, and thereby close to the free surface while in the 
flowing zone.
The walls are smooth (sw) or rough made of a monolayer 1, 2 or 4~mm particles.}
\label{driftcurve14}
\end{figure}

To characterize more precisely the dependence of the drift and the curvature
on the roughness of the wall, both are measured along the length of a 0.14~m 
long tumbler
with wall roughnesses varying from smooth wall (sw) to 4~mm rough walls (cylindrical wall and endwalls made of
4~mm particles). 
The drift is strongly affected by
the roughness of the wall (Fig.~\ref{driftcurve14}a). Each
$x$-axis crossing corresponds to the boundary between 2 recirculation cells with the middle $x$-axis crossing corresponding to the symmetry plane at $x=0$.
The axial length of the endwall cell decreases with increasing roughness of the wall, 
while the central cell axial length increases. 
The maximum amplitude of the drift is approximately proportional the size of the recirculation cell. 

The curvature of the trajectory, 
measured as the maximum axial displacement  
 only slightly increases 
with the roughness of the walls (Fig.~\ref{driftcurve14}b). This is different from a spherical tumbler where both
drift and curvature are strongly dependent on the wall roughness \cite{DOrtonaThomas15}. 



\begin{figure}[htbp]
\includegraphics[width=0.95\linewidth]{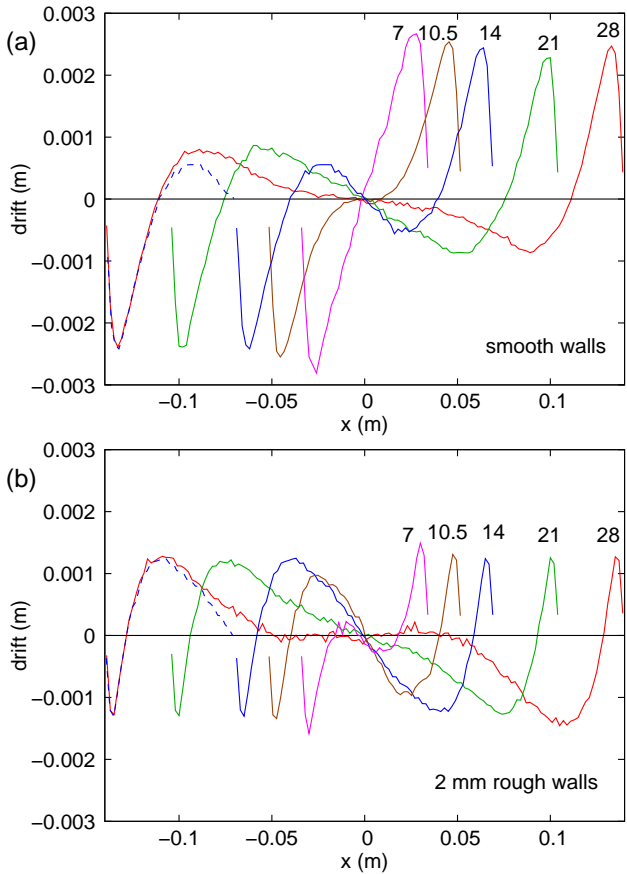}
\caption{Axial drift near the surface measured along tumblers of increasing length ($L=0.07$,
0.105, 0.14, 0.21 and 0.28~m, labeled in cm in the figure{\color{black}, corresponding to $L/2R=0.5$, 0.75, 1, 1.5, 2)}) with a fill level of 30\%. The tumblers have (a) smooth
walls or (b) 2 mm rough walls and rotate at 15 rpm {\color{black}($Fr= 0.018$)}. In the left part of both
figures, the
corresponding drift in 0.14~m long tumbler (dashed blue) after shifting
it to the left wall is shown for comparison.}
\label{driftvslength}
\end{figure}
Both the length of the tumbler and the wall roughness impact the axial drift and the size of the recirculation cells
(Fig.~\ref{driftvslength}).
For either wall roughness, the magnitude of the drift and the size of
the endwall cells are similar regardless of the tumbler
length (endwall cell size is the distance from the endwall to the first zero value of the drift). However rough walls result in reduced 
 drift for the endwall
 cells and larger axial drift for the central
 cells compared to smooth walls. 
The endwall cells are also surprisingly shorter in length for the rough walls compared
to smooth walls.  

For both roughnesses, the longest tumbler ($L=0.28$~m) has a cell at each
endwall, with an adjacent counter rotative central cell, and a zone having negligible axial drift at the center of the
tumbler  (Fig.~\ref{driftvslength}).  
We have considered even longer tumblers
($L=0.35$ and 0.42~m), but no additional cells appear near the center
of the tumbler. Instead, the central zone with no 
drift widens as the tumbler length increases and the central counter-rotating cells in Fig.~\ref{figintro} remain next to the
endwall cells, with a nearly constant size as the tumbler length increases. 
For all long tumblers, the endwall cells are similar, independent
of the tumbler length. This is made evident by shifting the axial drift
plot for the 0.14~m long tumbler (blue dashed curve) toward the left endwall 
so it overlays the 0.28~m long tumbler drift plot. The two curves perfectly coincide from the endwalls to the 
first crossing with the $x$-axis and, in the case of the rough wall,
even further toward the center of the tumbler. 
Thus, for very long tumblers, the character of the central cells is
independent of the tumbler length.

As the length of the tumbler decreases, the central cell is affected more than the endwall cell.  First,
the central zone with no drift disappears (Fig.~\ref{driftvslength}). Then, the size of the 
central cells decreases and, for very short tumblers, the drift
amplitude in the central cell decreases and its size is
constrained. The central
recirculation cell disappears altogether for the 0.07~m and 0.105~m long
smooth tumblers, while it persists for rough walls, though with substantially decreased drift and shorter length. 
For the 0.105~m long smooth tumbler, there is a short region with no
drift around the center of the tumbler, corresponding to an intermediate case
between 2 cells (as for
0.07~m long) and 4 cells (as for 0.14~m long).

\begin{figure}[htbp]
\includegraphics[width=0.95\linewidth]{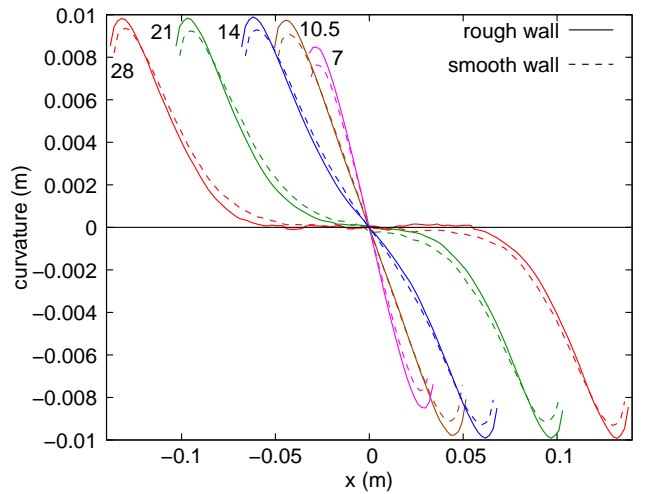}
\caption{Trajectory curvature for  tumblers of increasing length 
($L=0.07$ to 0.28~m{\color{black}, $L/2R=0.5$ to 2}) with a fill level of 30\% and rotating at
15~rpm {\color{black}($Fr=0.018$)} with a smooth wall (dashed curves) and 2~mm rough wall (solid curves).}
\label{curvelong}
\end{figure}

The curvature of the trajectories is similar between
smooth and rough walls, though
it is slightly higher for rough walls (Fig.~\ref{curvelong}). 
It is generally understood that the trajectory curvature is induced by the endwall friction \cite{PohlmanOttino06,SantomasoOlivi04,ChenOttino08,ManevalHill05,PohlmanMeier06,DOrtonaThomas18}. 
Endwall friction reduces the
flux of particles along the endwall. To accommodate this flux reduction, flowing particles must curve away
from the endwall in the upper part of their trajectory and
back toward the endwall in the lower part.
\begin{figure}[htbp]
\includegraphics[width=0.95\linewidth]{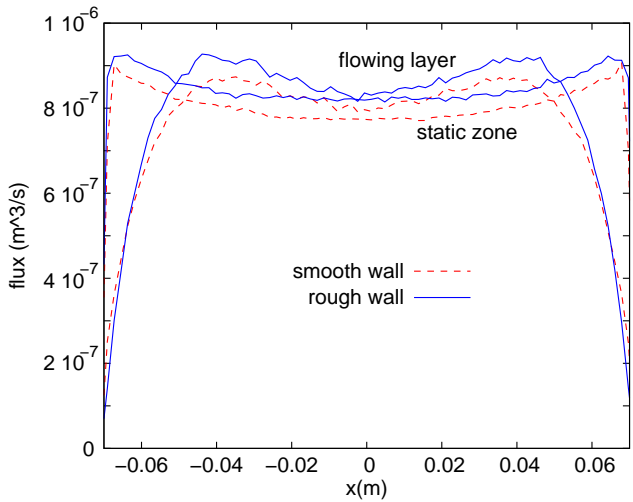}
\caption{Profile of the streamwise flux of material in a tumbler with rough (solid)
and smooth (dashed) walls. The flux is measured either in the static zone
or in the flowing layer.
The tumbler is 0.14~m {\color{black}($L/2R=1$)} long, filled to 30\%, and rotates at 15~rpm {\color{black}($Fr=0.018$)}.}
\label{flux}
\end{figure}
In light of this mechanism, what is surprising is that a much rougher
endwall (2~mm rough wall) only increases the curvature slightly. To 
investigate this further, Fig.~\ref{flux} compares the
streamwise flux in the flowing layer to that in the static zone in solid 
body rotation. The flux (and streamwise velocity) profiles have two local maxima that seem to be related to the existence of 4 cells. Note that short tumblers with no central cells have a velocity profile with a single maximum, although the tumbler length for the transition between 2 and 4 cells does not exactly coincide with the transition between one and two maxima (see Fig.~S1 in Supplemental Material \cite{Supplemental}). With rough walls,
the flux in the flowing layer is reduced near the endwalls and increased at about
0.03~m away from the endwalls. At the same time, the flux in the solid body rotation zone 
increases near the endwalls, as the thickness of
the flowing zone decreases (as is evident in Fig.~\ref{displacementmap}, discussed shortly). 
To adapt to the flux difference, particles have to follow curved trajectories. 
Thus, the curvature increases, as the flux difference increases, but only slightly because
 fluxes in the static zone and in the flowing layer are quite 
similar for rough and smooth endwalls (Fig.~\ref{flux}).
This result is different from the case of a rotating spherical tumbler in which
the curvature substantially increases with the wall roughness (by a factor of 3). Of course,
in a sphere, the fluxes are influenced both by the wall friction and the wall geometry \cite{DOrtonaThomas15}.

\subsection{Geometry of the recirculation cells}

\begin{figure}[htbp]
\includegraphics[width=0.99\linewidth]{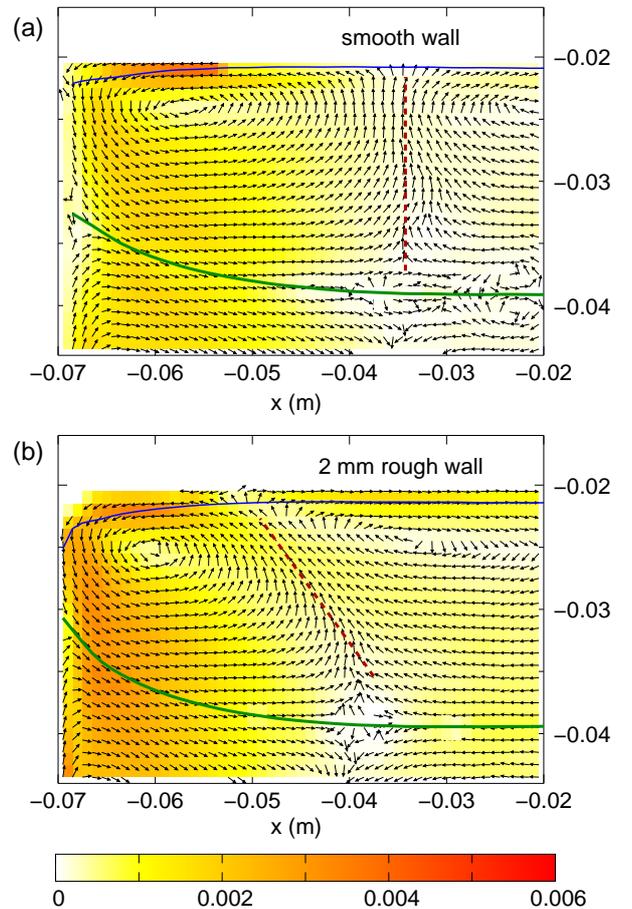}
\caption{Displacement maps for (a) a smooth wall tumbler and (b) a 2~mm rough 
wall tumbler that is 0.14 m {\color{black}($L/2R=1$)} long, filled to 30\%, and rotates at 15 rpm {\color{black}($Fr=0.018$)}. The dashed red lines indicate the boundary between the 
endwall recirculation cell and the central recirculation cell. The upper horizontal curve (blue) shows the
free surface  based on a volume concentration of 0.3, and the bottom horizontal curve (green) indicates the
lower boundary of the flowing layer based on a null velocity in the laboratory reference frame.
All vectors have the same length. The color map gives 
the displacement amplitude in meters.}
\label{displacementmap}
\end{figure}

Further information about the nature of the recirculation cells is gained from a vector map of particle displacement
between
two successive passes through a plane perpendicular to the free
surface and including the rotation axis (Fig.~\ref{displacementmap}). Particle mean trajectories start from this plane, make a complete circuit through the
flowing layer and static zone and then cross the plane again. Each arrow 
in the displacement vector map is drawn to show the direction (but not
magnitude) of the displacement 
between the starting and the ending position. The color intensity indicates the magnitude of the displacement.
Only the flowing layer is shown because the displacement vector map of a recirculation cell in the solid body rotation zone is simply a stretched mirror of that 
cell in the flowing layer (see Fig.~S2 in Supplemental Material for complete 
displacement maps \cite{Supplemental}). 

Regardless of wall roughness, the displacement map is 
consistent with the endwall cell trajectory in Fig.~\ref{figintro}, 
with a surface drift toward the endwall
and a drift deeper in the flowing layer toward the center of the tumbler.
The boundary between endwall and central cells is vertical for smooth walls but
tilted for rough walls. This tilt significantly reduces the size of
the endwall cell at the surface of the flowing layer, consistent
with the decrease of the endwall cell length measured near the surface
(Fig.~\ref{driftcurve14}(a)). The thickness of the flowing 
layer very near the endwall decreases with 
increasing wall roughness (the thickness is taken between the blue curve indicating the top of the 
flowing layer based on volume concentration and the green curve indicating the bottom based on a null velocity). Furthermore, the displacement amplitude is greater near the endwall
 for the rough than for the smooth endwalls. 
This comes about because the thickness of the static zone
in solid body rotation (below the green curve in Fig.~\ref{displacementmap}) 
is increased near the endwall, especially for rough walls, with the consequence
of a higher flux in the solid body rotation near the endwall. 
 Although it is not shown here, the dynamic angles of repose are nearly
identical for both smooth and rough walls in spite of the differences 
in the two displacement maps. 


\subsection{Rotation speed}            
\label{vitesse}

\begin{figure}[htbp]
\includegraphics[width=0.95\linewidth]{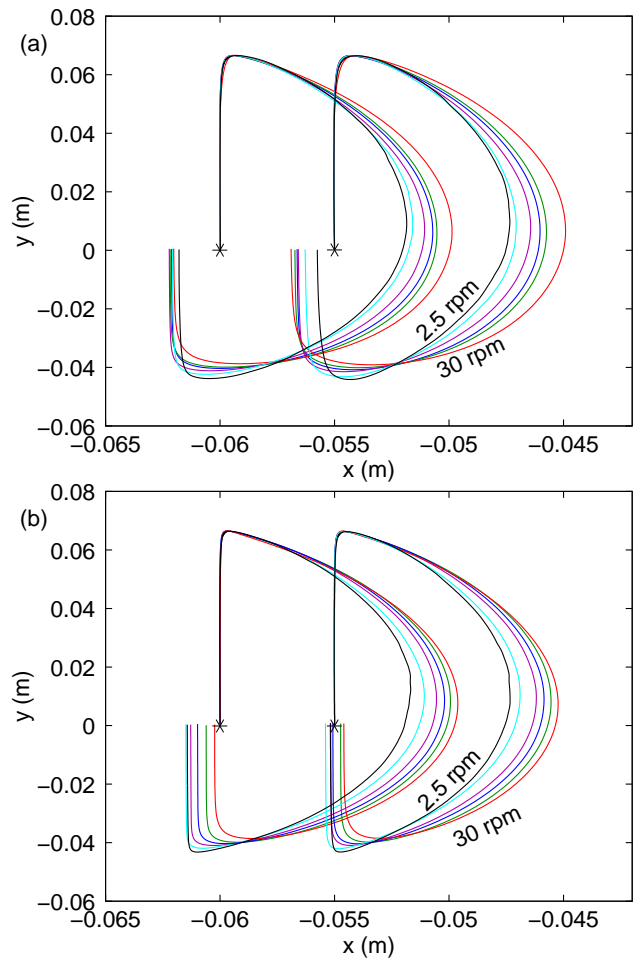}
\caption{Mean particle trajectories (top view) starting from
two initial positions (marked by a star), close to the left endwall,
and for 6 different rotation speeds (2.5, 5, 10, 15, 20 and 30 rpm{\color{black}, corresponding to $Fr=0.0005$ to 0.070}).
The 0.14~m {\color{black}($L/2R=1$)} long tumblers have (a) smooth walls or (b) 2~mm rough walls.
The fill level is 30\%.}
\label{cyltrajvsvit}
\end{figure}
Rotation speed is an important parameter in tumbler flows. For
bidisperse flow in cylindrical tumblers, below some velocity threshold, axial
segregation may disappear \cite{HillKakalios94,HillKakalios95}.
In spherical tumblers, increasing the
rotation speed may (depending on the fill level) induce a transition in 
the segregation patterns 
\cite{ChenLueptow09,DOrtonaThomas16} as a result of
the curvature of the particle
trajectories. For monodisperse flows in a spherical tumbler, increasing the
rotation speed increases the trajectory
curvature \cite{DOrtonaThomas15,DOrtonaThomas16}.

Figure~\ref{cyltrajvsvit} shows mean trajectories of particles starting at two initial locations
near the left endwall of a 14~cm long tumbler and for rotation
speeds ranging from 2.5 to 30~rpm. For both tumbler roughnesses and 
both initial positions, the trajectory curvature increases with increasing rotation speed.
In the case of a smooth wall tumbler, the drift increases with 
increasing rotation speed while it decreases in the case of a 2~mm 
rough tumbler. 

\begin{figure}[htbp]
\includegraphics[width=0.95\linewidth]{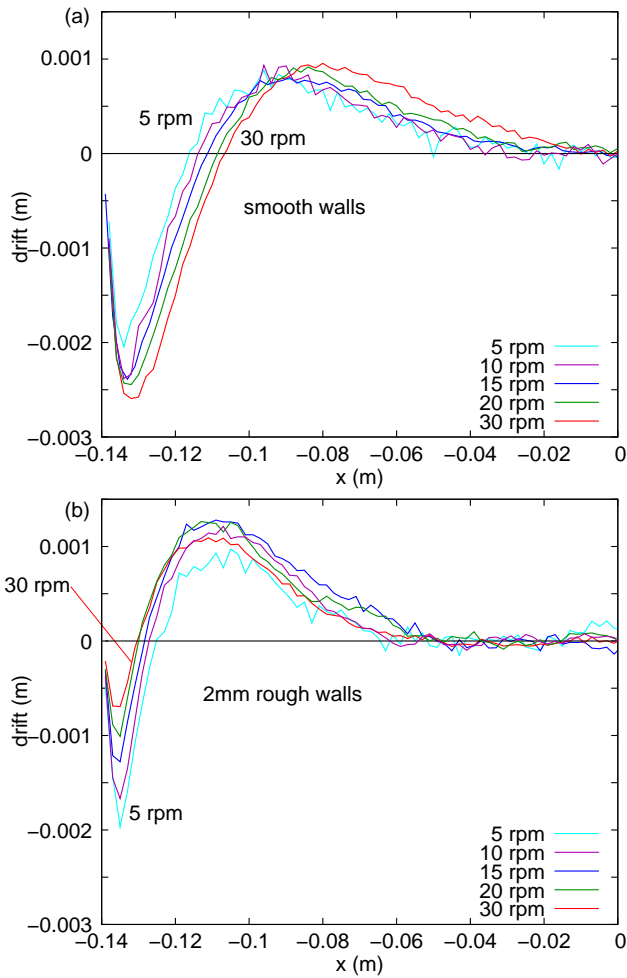}
\caption{Drift measured along the length of the left half of a 28~cm {\color{black}($L/2R=2$)} tumbler for various
rotation speeds (5 to 30 rpm{\color{black}, $Fr=0.002$ to 0.070}).
The tumbler has (a) smooth walls or (b) 2~mm rough walls and is filled at 30\%.
See Fig.~S3 in Supplemental Material for 14~cm long tumblers \cite{Supplemental}.}
\label{driftvsvit}
\end{figure}

Considering the drift along the entire length of a 28~cm long tumbler (to avoid any influence of the tumbler length on the results) in Fig.~\ref{driftvsvit} demonstrates a consistent change with rotation speed for the smooth wall tumbler, with an increase of the endwall cell size with increasing rotation speed. 
The rough wall tumbler displays slightly less variation with rotation speed, though there
is an opposite tendency to that of the smooth tumbler, as discussed in section \ref{link}. Results for 14~cm long tumblers are very similar to
that shown in Fig.~\ref{driftvsvit} for 28~cm long tumblers, except that there is no central zone with no drift and the central cells
are slightly reduced in size due to the symmetry plane in the center of the tumbler
($x=0$) for the smooth case (see Fig.~S3 in Supplemental Material \cite{Supplemental}).

\begin{figure}[htbp]
\includegraphics[width=0.95\linewidth]{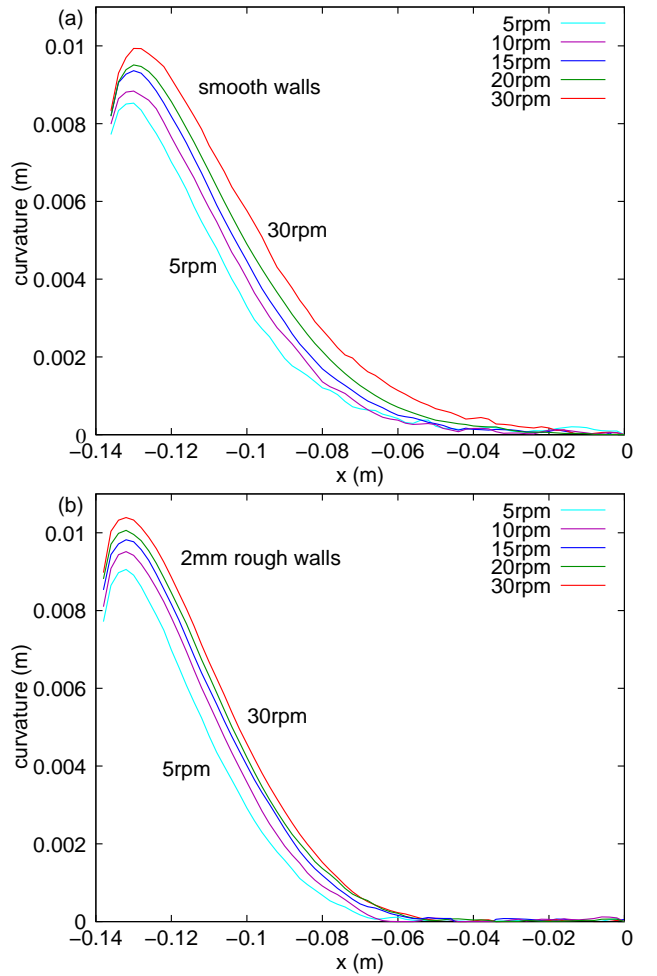}
\caption{Curvature measured along the 28~cm {\color{black}($L/2R=2$)} tumbler for various
rotation speed (5 to 30 rpm{\color{black}, $Fr=0.002$ to 0.070}).
The tumbler has (a) smooth walls or (b) 2~mm rough walls and is filled at 30\%. Curvature for only the left half of the tumbler
is shown. See Fig.~S4 in Supplemental Material for 14~cm long tumblers \cite{Supplemental}.}
\label{curvevsvit}
\end{figure}
The curvature is also affected by the rotation speed (Fig.~\ref{curvevsvit}).  
For both roughnesses, increasing the rotation speed increases the curvature everywhere
in the tumbler. As observed in Figs.~\ref{driftcurve14}(b) and \ref{curvelong}, the 
curvature decreases more rapidly  moving toward the center $x=0$ 
for the rough tumbler cases.

For brevity, the effect of rotation speed on displacement maps is only discussed here with reference to Supplemental Material Figs.~S5 and S6  \cite{Supplemental}. For both smooth and
2~mm rough tumblers,  the displacement maps are similar to those in Fig.~\ref{displacementmap} except for the dependence of the boundary between the endwall recirculation cell and the central recirculation cell. Increasing the rotation speed tilts the
boundary (dashed thick lines in Fig.~\ref{displacementmap}) progressively further clockwise.
In the smooth wall tumbler (and left half-tumbler), 
the tilt is leftward for lower rotation speeds and rightward for higher speeds
than that shown in Fig.~\ref{displacementmap}. Thus, increasing the rotation
speed induces the enlargement of the endwall cell at the free surface,
consistent with Fig.~\ref{driftvsvit}. In the rough wall 
case, the tilt is slightly more leftward for lower rotation speeds and slightly closer to vertical for higher rotation speeds although it never becomes vertical even at 30 rpm.  In addition, the tilt is accompanied by a small displacement of the boundary toward
the endwall thereby reducing the endwall cell size at the free surface.  
 
Far from the endwall, drift and curvature seem directly 
linked (Figs.~\ref{driftvsvit} and \ref{curvevsvit}). In the rough wall case, 
the drift and curvature vanish around $x\simeq -0.05$~m. For the
smooth wall tumbler, they vanish closer to the middle of the tumbler, between $x=-0.04$~m and -0.02~m.

%



\subsection{Fill level}                 

Finally, we consider the effect of the fill level.
In a spherical tumbler, the
curvature of trajectories is increased by a factor of 2
in a smooth sphere and by 50\% in a rough sphere when reducing the fill
level from 50\% to 25\% and the axial drift 
is also strongly modified \cite{DOrtonaThomas15}. Furthermore, the
segregation pattern for bidisperse particles in a spherical tumbler is 
completely reversed by changing the fill level
\cite{ChenLueptow09,DOrtonaThomas16}. 

\begin{figure}[htbp]
\vspace{2mm}
\includegraphics[width=0.95\linewidth]{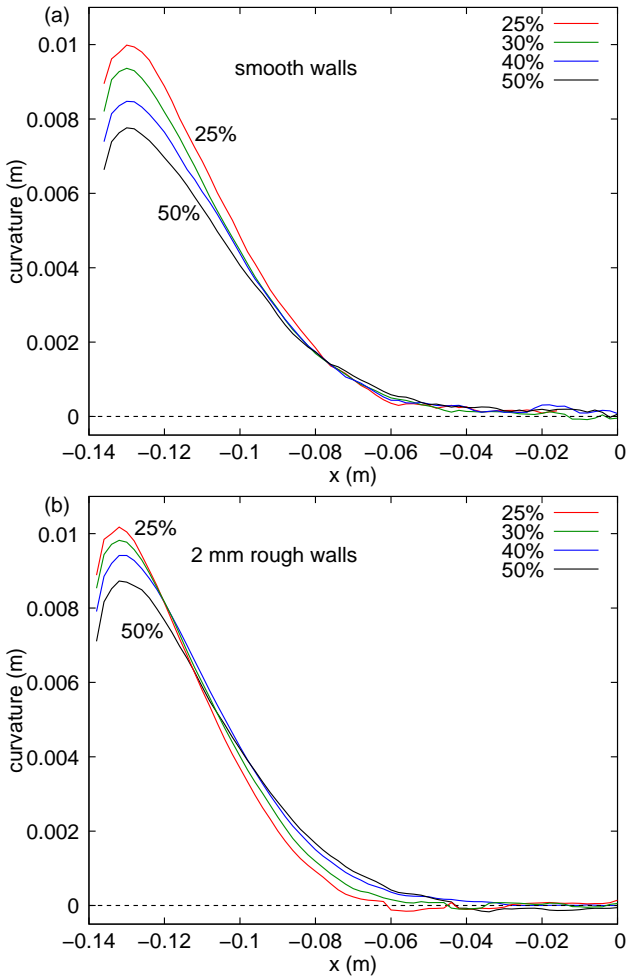}
\caption{Trajectory curvature measured along 28~cm {\color{black}($L/2R=2$)} tumbler for various fill levels ranging from 25\% to 50\%. The tumbler has (a) smooth or (b) 2~mm rough walls and rotates at 15~rpm {\color{black}($Fr=0.018$)}. Only the left half of the tumbler is shown.}
\label{curvefill}
\end{figure}
\begin{figure}[htbp]
\vspace{2mm}
\includegraphics[width=0.95\linewidth]{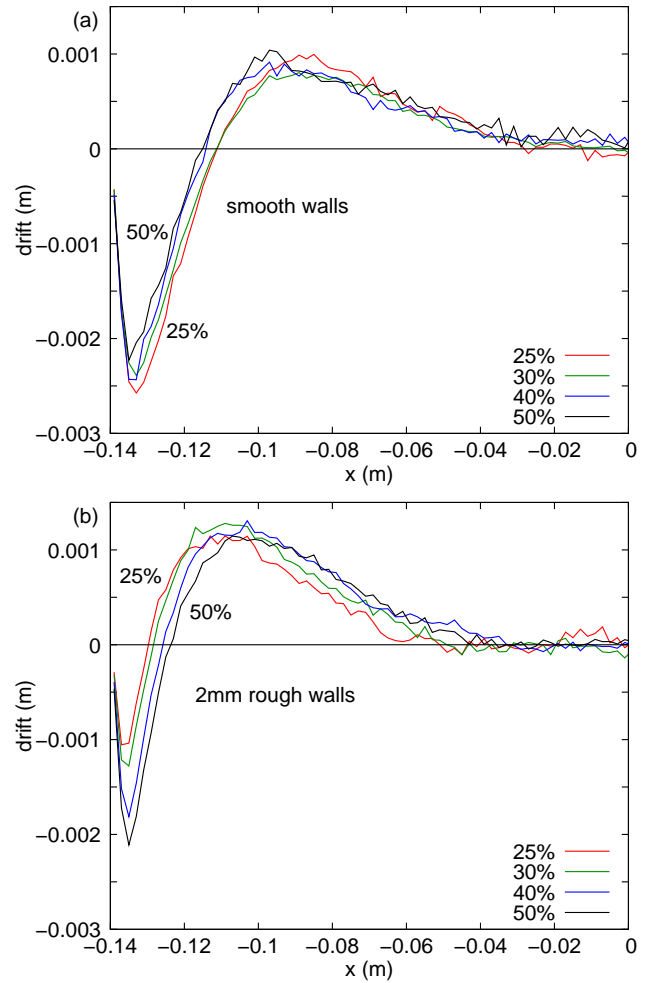}
\caption{Drift measured along 28~cm tumbler {\color{black}($L/2R=2$)} for various
fill levels ranging from 25\% to 50\%.
The tumbler has (a) smooth or (b) 2~mm rough walls and rotates at
15~rpm {\color{black}($Fr=0.018$)}. Only the left half of the tumbler is shown.}
\label{driftvsfill}
\end{figure}
In the cylindrical tumbler, the fill level also modifies the particle trajectories, 
but in a much weaker way. 
In a smooth tumbler, the maximum curvature is reduced
by nearly 25\% near the endwall when increasing  the fill level from 25\% to 50\% (Fig.~\ref{curvefill}).
The reduction in curvature is somewhat smaller in a rough tumbler.
Along the length of the tumbler, the dependence of the curvature on the fill level is more complex. For the rough wall case in Fig.~\ref{curvefill}(b), the curvature curves cross each other so that in the region (-~0.11~m~$<x<-0.04$~m), the largest curvature is associated with the greatest fill level, opposite that near the endwall. For the smooth wall case,  the curvature curves 
simply merge moving closer to the center of the tumbler [Fig.~\ref{curvefill}(a)].

For a smooth tumbler, 
increasing the fill level has little effect on the drift, except to slightly shorten the 
endwall cell and slightly reduce the magnitude of the drift (Fig.~\ref{driftvsfill}). 
By contrast, for a 2-mm rough tumbler, the drift varies more and in the opposite way:   
the endwall cell increases in size with the fill level. The differing dependence on fill level will be discussed in section \ref{link}. Far from endwall ($x\gtrsim-0.08$~m), drift and curvature vary with $x$ in a similar way for smooth and rough walls.
Similar results for the curvature and drift are obtained for 14~cm long 
tumblers (Supplemental Material, Figs.~S7 to S9 \cite{Supplemental}). In addition, the boundary between cells in the displacement maps rotate in opposite directions with increasing fill level, reflecting the different dependence of the cells themselves on fill level for smooth and rough walls (Supplemental Material, Figs.~S10 and S11  \cite{Supplemental}).

\section{Endwall versus cylindrical wall roughness}

Clearly, wall roughness has a surprisingly strong effect on the flow in a cylindrical tumbler. To better understand the relative effects of the roughness of the
endwall and cylindrical wall, the roughness
of each wall can be modified independently. Several studies
have modified the sense of rotation
or the friction of the tumbler endwalls,
but the cylindrical wall remains smooth in all cases \cite{ChenOttino08,ChenLueptow11,HuangLiu13}.
Here four combinations of cylindrical wall and endwall roughnesses are considered in order to tease out how wall roughness affects the flow.

\subsection{Mixed wall roughness}

\begin{figure}[htbp]
\includegraphics[width=0.95\linewidth]{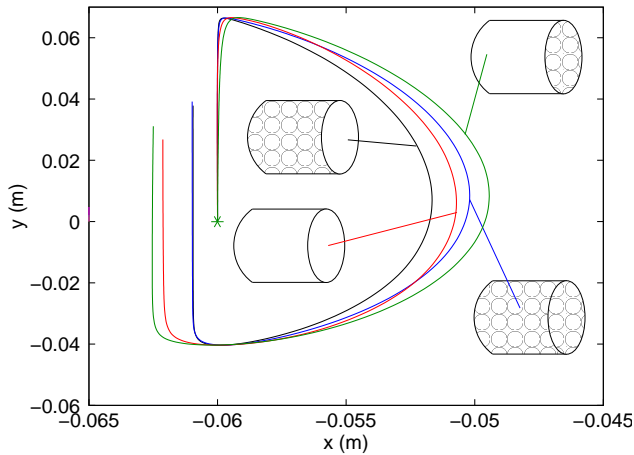}
\caption{Mean trajectories of particles (top view) starting
from an initial position (marked by a star) 3~mm from the cylindrical wall, 
close to the left
cylindrical wall, for 4 different combinations of wall roughnesses: smooth (red), 2~mm rough
(blue), mixed smooth cylindrical and rough endwalls (green), and mixed rough
cylindrical and smooth endwalls (black).
The tumbler is 0.14~m {\color{black}($L/2R=1$)} long and rotates
at 15 rpm {\color{black}($Fr=0.018$)}. The $x$ axis is stretched
compared to the $y$ axis.}
\label{trajmixed}
\end{figure}

Figure~\ref{trajmixed} shows four typical trajectories starting from the same
location in tumblers with the four possible combinations of wall roughnesses. Several differences are immediately evident.
First, endwall roughness favors a large curvature whereas cylindrical wall roughness decreases curvature. Second, the drift is mainly
controlled by the cylindrical wall roughness. In the case of a rough cylindrical wall, trajectories differ depending on the endwall roughness, but finish with exactly
the same drift.  A similar result occurs for the smooth cylindrical wall, where
the axial drifts are nearly the same, but are
noticeably larger than those for the rough cylindrical wall.

 \begin{figure}[htbp]
\includegraphics[width=0.95\linewidth]{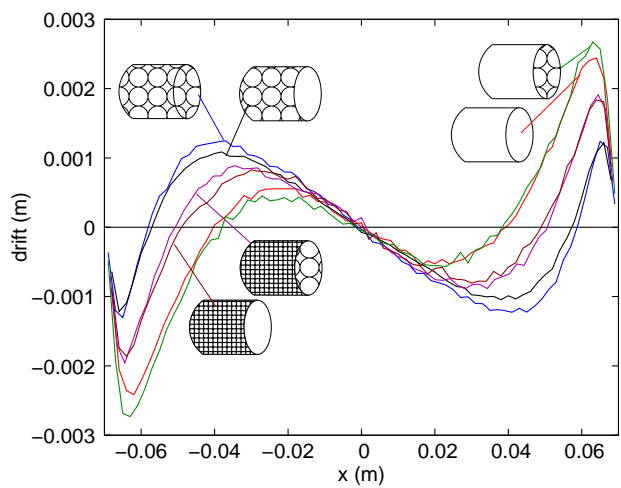}
\caption{Drift measured along 0.14~m {\color{black}($L/2R=1$)} long tumblers
for 6 different wall roughness combinations of smooth, 0.75~mm rough, and 2~mm
rough walls. The tumblers rotate at 15 rpm {\color{black}($Fr=0.018$)} and are filled to 30\%.}
\label{driftmixed}
\end{figure}

The drift along the entire length of the tumbler is 
nearly identical for the same cylindrical wall roughness regardless of the 
endwall roughness (Fig.~\ref{driftmixed}). 
The drift for an intermediate roughness
(cylindrical wall made of 0.75 mm particles) 
falls between the smooth and 2~mm rough wall case.
 Consistent with Fig.~\ref{driftcurve14}, the amplitude of the drift 
decreases in the endwall cell with increased cylindrical wall roughness but increases in the central cell.

\begin{figure}[htbp]
\includegraphics[width=0.93\linewidth]{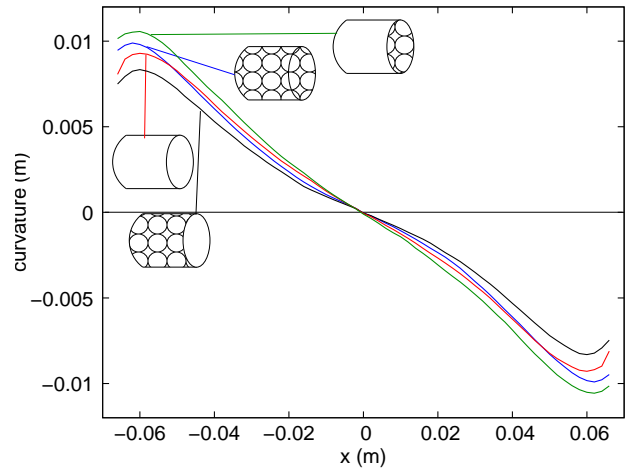}
\caption{Curvature measured along 0.14~m {\color{black}($L/2R=1$)} long tumblers for 4 combinations of wall roughnesses: smooth (red), 2 mm rough (blue),
2 mm rough endwalls and smooth cylindrical wall (green) and smooth endwalls 
and 2 mm rough cylindrical wall (black). The tumblers rotate at
15 rpm {\color{black}($Fr=0.018$)} and are filled to 30\%.}
\label{curvaturehyb}
\end{figure}

On the other hand, the curvature of particle trajectories differs for the four wall roughness combinations shown in Fig.~\ref{curvaturehyb}.   In fact, the roughness of the endwalls and the cylindrical wall have opposite effects. 
Rough endwalls and smooth cylindrical walls result in more curvature.
Near the endwalls the greatest curvatures are obtained with rough endwalls, consistent with Fig.~\ref{driftcurve14}(b). Moving toward the center, rough cylindrical walls induce
a slightly larger decrease of the curvature.
As we will show later, the role of the endwall roughness is to
induce trajectory curvature
near the endwall, while the role of the 
cylindrical wall roughness is to reduce the curvature moving away from the endwalls. 
The curves for
0.75~mm cylindrical walls are not shown in Fig.~\ref{curvaturehyb}, but they fall
between the smooth and 2~mm rough wall cases as expected.


\begin{figure}[htbp]
\includegraphics[width=0.95\linewidth]{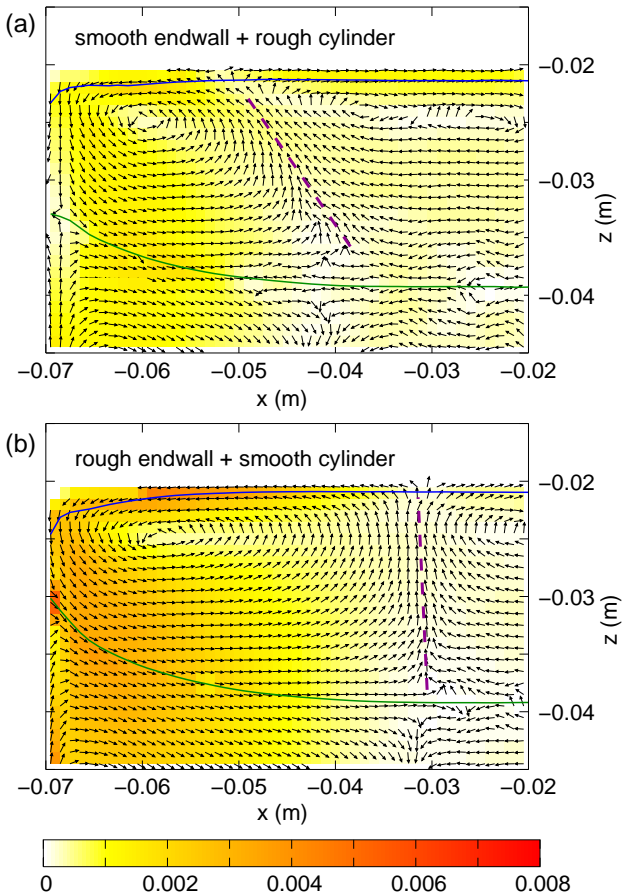}
\caption{Displacement map for tumblers (a) with smooth endwalls and a 
2~mm-rough cylindrical wall and (b) 2~mm-rough endwalls and a smooth cylindrical
wall. {\color{black} The tumblers are 0.14 m ($L/2R=1$) long, filled to 30\%, and rotate at 15 rpm ($Fr=0.018$)}. The dashed lines indicates the boundary between the endwall 
 cell and the central cell. All vectors have the same length. The color map gives the displacement map amplitude in meters and colors are 
identical to Fig.~\ref{displacementmap} and have been
completed up to 0.008~m with a darker red.}
\label{maphybbyhsurf}
\end{figure}

Comparing the displacement maps for the four combinations 
(Figs.~\ref{displacementmap}
and~\ref{maphybbyhsurf}) provides additional information about the character of the recirculation cells in these cases. The shape of the recirculation cell clearly depends almost entirely on the cylindrical wall roughness, as indicated by
the orientation of the boundary between cells (thick
dashed lines), which is vertical for a smooth cylindrical wall and angled for a rough cylindrical wall, regardless of the endwall roughness.
The cylindrical wall roughness controls the cells through the thickness of the flowing layer, not just at the surface. However, the amplitude of the vertical displacement near the endwall, only depends on endwall roughness (visible as the color near the left endwall in Figs.~\ref{displacementmap} and \ref{maphybbyhsurf}).

\subsection{Connecting wall roughness and drift}       
\label{link}

\begin{figure}[htbp]
\includegraphics[width=0.95\linewidth]{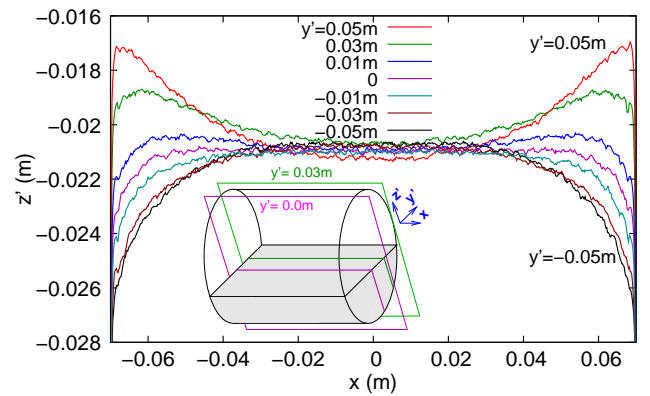}
\caption{Free surface along a 14~cm {\color{black}($L/2R=1$)} long rough tumbler in $x-z'$ planes perpendicular to the free surface. The tumbler is filled at 30\% and rotates at 15~rpm {\color{black}($Fr=0.018$)}.}
\label{freesurfmix}
\end{figure}

It is quite clear that the cylindrical wall and endwall roughnesses alter the particle trajectories, but the question is how these are linked.  The answer appears to lie in the topography of the free surface. 
Figure~\ref{freesurfmix} shows
sections of the free surface in several $x-z'$ planes perpendicular to the free surface, 
where $(x,y',z')$ is a reference frame tilted to be aligned with the free
surface at the center of the tumbler at $x=0$.
The angles at which the reference frame is tilted for the four roughness
configurations differ by less than 0.4 degrees.
To provide context for the subsequent discussion, the surface topography at
several $x-z'$ sections is shown for a 14 cm long rough tumbler in
Fig.~\ref{freesurfmix} as an example. 
 Most notable are bumps in the topography near the endwalls for upstream locations ($y'>0$). Further note that even though the profile at $y'=0.01$~m presents small bumps, the depressions at the endwalls have a larger amplitude. Further downstream, the bumps do not occur at all and instead the surface level  simply decreases near the endwall. When comparing corresponding curves (for example $y'=-0.03$~m and $y'=0.03$~m), it is evident that near the endwall ($x\lesssim -0.06$~m) each depression has a larger amplitude than its corresponding bump. Both the upstream surface bumps and the downstream surface depressions diminish moving away from the endwall. {\color{black} This topography comes about due to friction at the endwall. The same number of particles enter the flowing layer along the entire length of the tumbler, but particles near the endwall flow down the slope more slowly than particles far from the endwall \cite{ManevalHill05,PohlmanOttino06,PohlmanMeier06,DOrtonaThomas18}. As a result, particles pile up near the endwall in the upstream portion of the flowing layer before they can flow axially away from the endwall causing the slight bump. In the downstream portion of the flowing layer, particles near the endwall are depleted before they can flow back toward the endwall, resulting in the depression.}

\begin{figure}[htbp]
\includegraphics[width=0.95\linewidth]{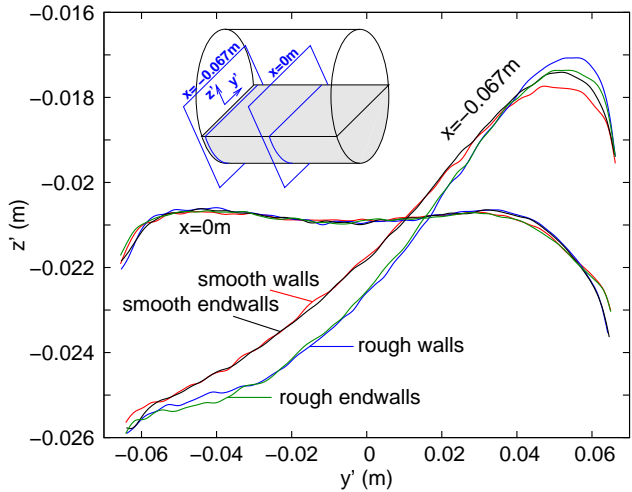}
\caption{Topography of the free surface at the center ($x=0$~m) and near
the left endwall ($x=-0.067$~m) in a $y'-z'$ plane where the $y'$ axis is
tilted such that it is parallel to the free surface at the center of the 
tumbler. The inset indicates the slicing planes. The tumblers are 0.14~m {\color{black}($L/2R=1$)} long 
and have 4 different wall roughness combinations: smooth (red), 2 mm rough (black),
2 mm rough endwalls and a smooth cylindrical wall (green), and smooth endwalls and a 2 mm rough cylindrical wall (black). The tumbler is filled at 30\% and rotates at 15 rpm {\color{black}($Fr=0.018$)}.}
\label{surflib30hyb}
\end{figure}

The difference in the topography of the free surface at the center of the tumbler and at the endwalls is shown more clearly in Fig.~\ref{surflib30hyb}, which presents $y'-z'$ cross sections of the free surface for the four different roughness conditions
at the center of the tumbler ($x=0$) and very near the left endwall
($x=-0.067$~m). 
In the tilted reference frame in Fig.~\ref{surflib30hyb} the surface topography is horizontal and nearly identical at the center of the tumbler for all four combinations of surface roughness.  However, near the endwall, the average angle of
free surface is steeper. {\color{black} The steep free surface near the endwall again comes about due to friction at the endwall. The bumps near the endwall correspond to the topography near the endwall being above that for $x=0$ in the upstream portion, and the depressions correspond to the topography being below that for $x=0$ in the downstream portion. The steepness is greater} for tumblers with rough
endwalls than for smooth endwalls.
As already observed in Fig.~\ref{freesurfmix}, near the endwall ($x=-0.067$~m), the depression at the bottom of the trajectory is larger than the bump at the top.
Axial variations of this topography induce axial displacements of particles: toward the tumbler center in the upper portion of the flowing layer since the surface is higher than at the center of the tumbler, and toward the endwall in the downstream portion of the flowing layer since the surface is lower at the endwall than at the center.  Since the difference between the surface heights at the endwalls and the center is smaller in the upstream portion of the flowing layer than in the downstream portion, there is 
a drift directed toward the endwall, which we denote $D_{ew}$. Consistent with the convention in Fig.~\ref{driftcurve14}, this drift is negative near the left endwall, positive near the right endwall, and negligible near the center for long tumblers.
Comparing the four tumbler roughness conditions in Fig.~\ref{surflib30hyb}, it is clear that
the topography curves superimpose for similar endwall roughnesses.  Thus, the
endwall roughness is primarily  responsible for the surface topography near the
endwall. Only a small
difference is induced by the cylindrical wall roughness and that is limited to  the very upper part of the flowing layer ($y'>0.04$~m in Fig.~\ref{surflib30hyb}).

To disentangle the effect of the endwall roughness and cylindrical wall
roughness on the trajectory drift, we hypothesize that the drift due to
the endwall $D_{ew}$ is linked to the topography variation along the length of
the tumbler.
To quantify the dependence of this topography on position along the length of the tumbler $x$, we fit the difference of the free surface heights 
taken at $y'=0.05$~m and $y'=-0.02$~m 
to a hyperbolic cosine $a\cosh(bx)+c$. This is most easily thought of as the difference between the curves for these two values of $y'$ in Fig.~\ref{freesurfmix}. The
variation of topography difference along the length of the tumbler, which we call the `topography
gradient,' is 
the $x$-derivative, $-a\,b\sinh(bx)$. 
We further assume that  $D_{ew}(x)$ has a variation with $x$ {\color{black} proportional} to that of this topography gradient.

We also assume that the cylindrical wall has a separate and additional effect on the drift
because it affects the axial displacement at the top and bottom of the flowing
layer. 
The impact of the cylindrical wall roughness on the drift is evident in 
Fig.~\ref{trajmixed} where the particle trajectories are 
directed further toward the endwall for a smooth cylindrical wall than for a rough wall.
 Cylindrical wall roughness opposes axial displacement in
the lower part of the trajectory because the trajectories are directed toward the cylindrical wall at a slight angle due to the trajectory curvature.
The effect of cylindrical wall roughness in the upper part of the trajectory is more subtle, 
but in the upper part of particles trajectories in Fig.~\ref{trajmixed}, the particles
move a small amount axially toward the center of the tumbler while still rising in the bed of particles for a smooth cylindrical wall, whereas, a rough cylindrical wall prevents any axial displacement before entering the flowing
layer. 
 
Although the effects of cylindrical wall roughness in the upper and lower portions of the flowing layer are opposite in direction, the stronger effect is at the lower portion of the trajectory. The combination of the upper and lower axial displacements creates a drift related to the cylindrical wall in the same direction as the curvature, i.e. toward the center of a cylindrical tumbler (e.g., compare axial drift for smooth and rough cylindrical walls but the same endwall roughness in Fig.~\ref{trajmixed} recalling that at this $x$ location the endwall drift is dominant).
We call this drift $D_{cyl}$, which is positive in the left part of the tumbler, and negative in the right part. 
$D_{cyl}$ acts opposite to the drift associated with the endwall friction $D_{ew}$.
Large cylindrical 
wall roughness favors a large $D_{cyl}$.
Since friction is enhanced by more tangential trajectories,
we assume that $D_{cyl}$ is proportional to the curvature with a coefficient $R$ depending on the cylindrical wall roughness. $D_{cyl}$ may also depends on other
 parameters including the rotation speed or the fill level. The value of this coefficient $R$ is unknown and needs to be estimated.

 With these definitions for $D_{ew}$ and $D_{cyl}$, we now assume that the total drift is $D=D_{cyl}+D_{ew}$.   
Near the endwall, $D_{ew}$ is dominant, and $D$ is directed toward the 
endwall. Moving toward the center of the tumbler, $D_{ew}$ decreases more 
rapidly than $D_{cyl}$, so that there is a position 
$x$ where $D$ is zero, which corresponds to the boundary between the 2 cells. Moving further toward the center, $D_{cyl}$
is dominant and $D$ is directed toward the center. 

\begin{figure}[htbp]
\includegraphics[width=\linewidth]{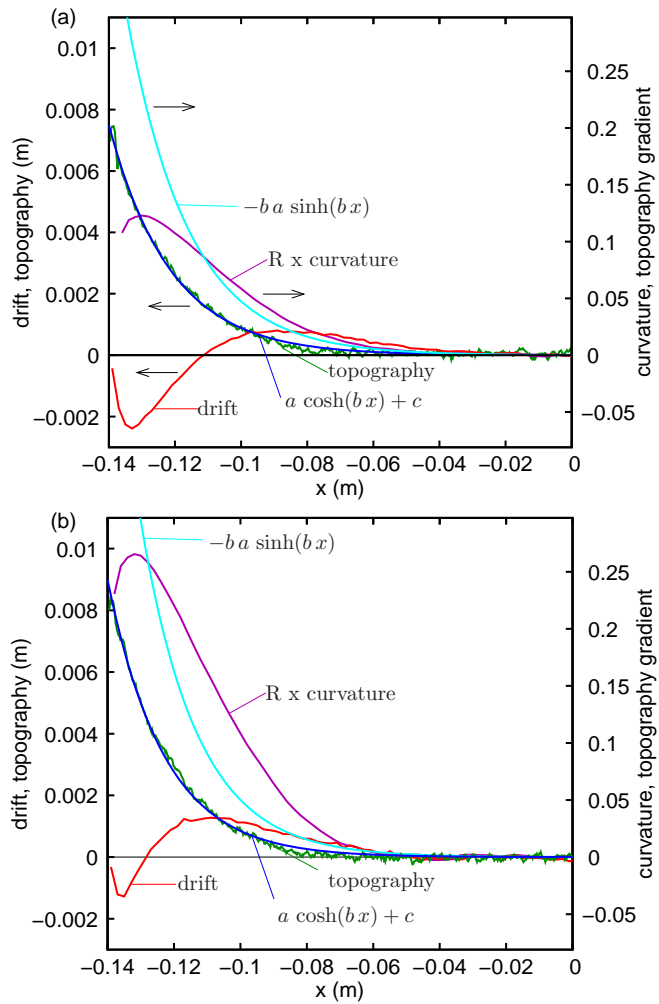}
\caption{Drift, free surface topography, its hyperbolic cosine fit (left
$y$-axis), $R\,\times$~curvature, and topography gradient (right $y$-axis) for 28~cm {\color{black}($L/2R=2$)} long 
tumblers with (a) smooth  and (b) 2~mm rough walls. The tumbler is filled at 30\% and rotates at 15~rpm {\color{black}($Fr=0.018$)}. Arrows indicate the vertical axis associated with the curve.}
\label{topocosh}
\end{figure}
Figure~\ref{topocosh} illustrates this approach in the case of 28~cm
long smooth and 2~mm-rough tumblers. A long tumbler
is preferred to avoid the influence of the central symmetry plane. The total drift, $D$, the free surface topography, and 
its hyperbolic cosine fit to the topography should be read on the left $y$-axis. On the right $y$-axis, we represent 
the measured curvature multiplied by the coefficient $R$ (adjustable){\color{black}, lumping together the two proportionality coefficients linking $D_{cyl}$ to curvature and linking $D_{ew}$ to topography. We also represent
the $x$-derivative of the topography difference 
(topography gradient) on the right $y$-axis. These two curves correspond to the evolution of
 $D_{cyl}$ and $-D_{ew}$, respectively.}
To obtain the
coefficient $R$, we assume that the endwall drift $D_{ew}$ exactly counterbalances 
the cylindrical wall drift $D_{cyl}$ at the $x$-location of the boundary between the endwall cell and the central cell. Thus, the two curves cross
at the location where $D=D_{cyl}+D_{ew}=0$, thereby setting the value for $R$. 

For the two cases shown in Fig.~\ref{topocosh}, the resulting value of $R$ is larger for the 2~mm rough wall tumbler ($R=27.0$) than for the smooth cylindrical wall tumbler ($R=13.1$). The value of $R$ is mainly linked to the cylindrical wall roughness, but, as we will show later, $R$ also depends on fill level and rotation speed. Thus, $R$ should be viewed as the combined effect of cylindrical wall roughness and the way particle trajectories interact with  the cylindrical wall (velocity, inclination, etc.). 

\begin{figure}[htbp]
\includegraphics[width=\linewidth]{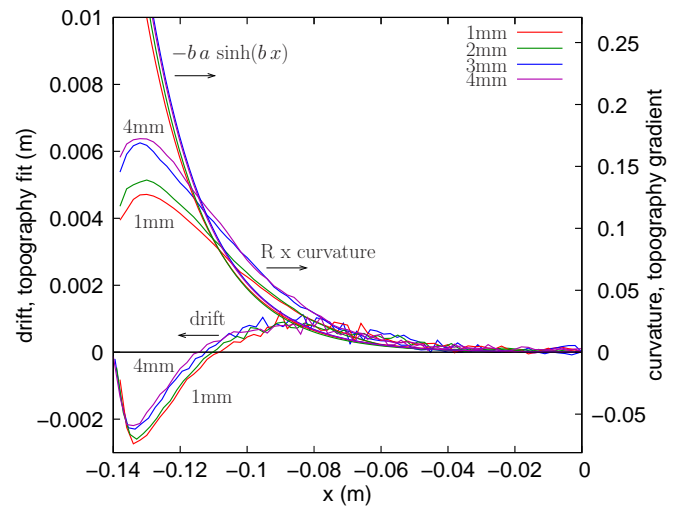}
\caption{Drift, topography fit, topography gradient, and $R$ times curvature  in the case
of 28~cm {\color{black}($L/2R=2$)} long tumblers with a smooth cylindrical wall and increasing endwall
roughnesses: 1, 2, 3, 4~mm. The tumbler is filled at 30\% and rotates at 15~rpm {\color{black}($Fr=0.018$)}. Arrows indicate the y-axis associated to the curve.}
\label{topoCourb}
\end{figure}
Based on the previous assumptions (and same rotation speed, fill level, etc.), mixed roughness tumblers having the same cylindrical wall roughness but
different endwall roughness should have the same value of the coefficient $R$. To demonstrate this,
Figure~\ref{topoCourb} shows $D$, $D_{cyl}$ ($R$ times curvature), and $D_{ew}$ (topography gradient)
for four tumblers having the same smooth cylindrical wall and endwalls 
with increasing roughness. The values of $R$ are 
13.6, 
13.1, 14.2 and 14.2 for the endwall roughness 1, 2, 3 and 4~mm, respectively. 
In addition, $R=12.9$ and 13.6 were measured for 0.5 and 1.5~mm rough
endwalls, respectively, while $R=13.1$ for a fully smooth tumbler.
The mean value is $R=13.5$ for a smooth cylindrical wall with the various endwall roughnesses. Since the relative standard deviation for the different endwall roughnesses is quite small (less than 4\%), $R$ can be considered independent of the endwall roughness, as assumed in the model.
\begin{figure}[htbp]
\includegraphics[width=\linewidth]{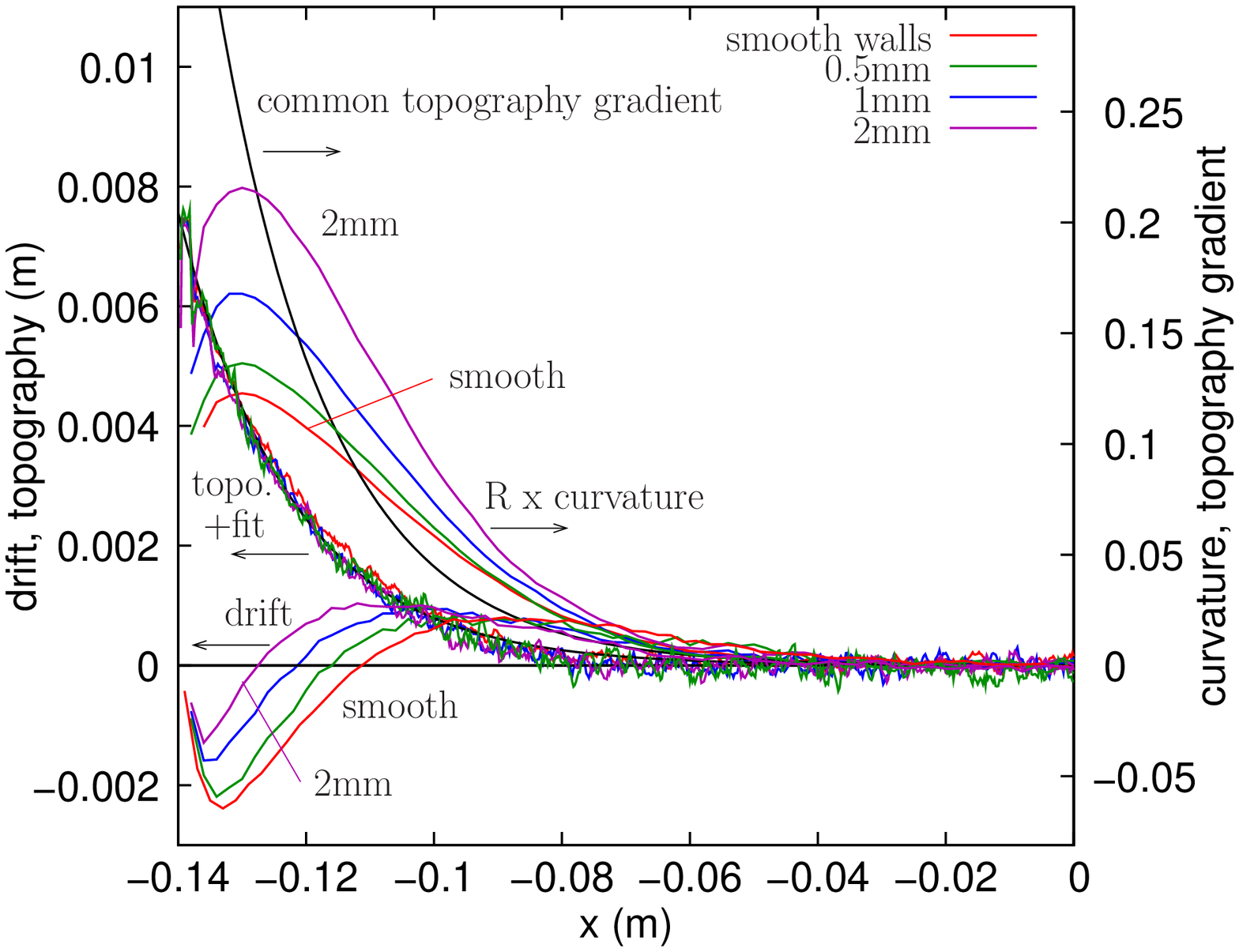}
\caption{Curves of drift, topographies and fits (labeled `topo.+fit'), topography gradient that is 
	common for all tumblers, and $R$ times curvature
in the case of long tumblers with smooth endwalls and increasing cylindrical wall
roughness: smooth, 0.5, 1, 2~mm. The tumbler is filled at 30\% and rotates at 15~rpm {\color{black}($Fr=0.018$)}. Arrows indicate the y-axis associated to the curve.}
\label{topobyh}
\end{figure}

In the reverse situation, tumblers with smooth endwalls and increasing
cylindrical wall roughness, we expect different values for $R$. 
As the topographies are nearly the same, a common fit can be used, leading to a single common topography
gradient curve representing the endwall drift $D_{ew}$  (Fig.~\ref{topobyh}).
Determining the values of $R$ so that intersections of $D_{ew}$ (topography gradient) with each  $D_{cyl}$ ($R$ times curvature) match the boundary between the endwall cell and the central cell yields increasing values of $R=13.2$, 15.2, 19.2 and 26.3 for increasing cylindrical 
wall roughnesses of smooth, 0.5, 1 and 2~mm, respectively. The value for $R$ 
and, accordingly, the cylindrical drift $D_{cyl}$ increase significantly with the 
cylindrical wall roughness, as expected. A plot of $R$ versus the cylindrical wall and
the endwall roughnesses is provided in Supplemental Material (Fig.~S12  \cite{Supplemental} ).

When tumblers are short enough, there
is only one cell in each half-tumbler, and the $R$ coefficient
cannot be estimated. Nevertheless, if $R$ values obtained
for smooth and rough longer tumblers are used, the topography gradient and R times curvature ($D_{ew}$ and $D_{cyl}$) curves do not cross, as would be expected. For smooth and rough tumbler lengths corresponding to the 4 cells to 2 cells transition, the two
curves are nearly tangent, also as expected (see Fig.~S13 in Supplemental
Material  \cite{Supplemental}).

There are some important points to make about the drift approach that we use to explain the effects of roughness of the tumblers. First, 
the model is robust since using the $R$ coefficient obtained for long tumblers shows that drift curves are
tangent or do not cross for shorter tumblers. Second, the approach is applicable near the boundary between the endwall cell and the central cell, but not over the entire length of the tumbler.  This is evident in Fig.~\ref{topocosh}, where the difference between the topography gradient and the curvature effect deviate substantially from the measured drift.  Hence, we do not expect this approach to provide
quantitative predictions of the dependence of the recirculation cell structure on wall roughness, 
rotation speed, fill level, and mixed wall roughness conditions considered in the next section. Nevertheless, it aides in understanding the underlying mechanism.

\subsection{Discussion}                   

Accounting for the drifts ($D_{ew}$ and $D_{cyl}$) induced by the endwalls and cylindrical wall separately, each of them dominating in one cell, provides a means to  
understand the mechanisms affecting the formation of the cells. It also decouples the recirculation in the central cell from the recirculation in the endwall cell, which is quite different from hydrodynamic systems where adjacent vortical cells are coupled at their interface. The proposed mechanism can explain why various configurations (larger drift in endwall cells or in central cells) are possible, even though the origin of the drift for all cells is the endwall friction.

For mixed roughness tumblers, increasing only the cylindrical wall roughness increases $R$ and consequently $D_{cyl}$ leading to a smaller endwall cell (Fig.~\ref{driftmixed}).
Increasing only the endwall roughness increases both the curvature (Figs.~\ref{curvaturehyb} and \ref{topobyh}) and the topography difference (Fig.~\ref{surflib30hyb}), thereby increasing both $D_{cyl}$
and $D_{ew}$ in a similar manner leaving the size of the cells almost unchanged (Figs.~\ref{driftmixed} and \ref{topoCourb}).
Increasing the roughness of both walls increases $D_{cyl}$ more
than $D_{ew}$, reducing the endwall cell size and the drift amplitude near the endwall. Indeed, in that case, the increase of $D_{cyl}$ has two contributions, one due to the curvature that is similar to the increase of $D_{ew}$ and another part due to the increase of $R$.

{\color{black}It is also important to note that the effects of endwall and cylindrical roughness, corresponding to $D_{ew}$ and $D_{cyl}$, respectively, are at the surface of the flowing layer {\color{black} and progressively decrease below the surface}.  After all, $D_{ew}$ is based directly on the surface topography gradient, and $D_{cyl}$ is based on the trajectory curvature resulting from interactions of {\color{black} the flowing particles at} the surface with the cylindrical wall, which is {\color{black} here measured for particles near the surface} of the flowing layer (recalling that {\color{black} the curvature measurement} is based on trajectories that start only 3 mm from the cylinder wall in the fixed bed and thereby remain near the surface of the flow as described in section \ref{link2}, {\color{black} noting that particle trajectories away from the free surface
also have some degree of curvature and drift). As a result, topography and roughness  weakly affect the deeper portion of the recirculation cells  with smaller values of $D_{ew}$ and $D_{cyl}$}.  This is evident in comparing Figs.~\ref{displacementmap}(a,b) and Figs. \ref{maphybbyhsurf}(a,b).  In both cases, changing from a smooth cylindrical wall to a rough one shifts the boundary between the recirculation cells from vertical to diagonal as a result of the position of the boundary at the surface moving toward the endwall, while the position of the boundary at the bottom of the flowing layer barely changes at all.} {\color{black} The largest effect on the bottom boundary results from the rotation speed, which shifts the position of the bottom
boundary toward the endwall for both roughnesses (see Fig.~S5 and S6 in Supplemental Material \cite{Supplemental}).
This demonstrates how several related effects define the spatial organization of the recirculation cells. 
A complete
description of trajectory drift and curvature requires further
modelling and is beyond the scope of this paper.}

Linking endwall drift to the topography gradient and cylindrical wall drift to the curvature allows us to consider the 
effects of rotation speed and fill level. For a smooth wall tumbler, increasing the rotation speed increases the size of the endwall cell and reduces the size of the central cell; the reverse occurs for a rough wall tumbler (Fig.~\ref{driftvsvit}). In both cases, increased rotation speed induces increased curvature (Fig.~\ref{curvevsvit}) and, hence, increased $D_{cyl}$. The topography difference also increases with rotation speed corresponding to an increase in $D_{ew}$. The difference between smooth and rough walls comes from the dependence of $R$ on rotation speed: $R$ increases only moderately with rotation speed for a smooth cylindrical wall (from $R=12.1$ to 14.4 for 5~rpm to 30~rpm) compared to a larger change for the rough case (from $R=20.9$ to 34.4 for 5~rpm to 30~rpm, see Figs. S14 and S15  \cite{Supplemental}). Consequently, the increase in $D_{cyl}$ dominates the increase in $D_{ew}$ for rough walls, so the endwall cell shrinks with increasing rotation speed, but the opposite occurs for the smooth tumbler. The strong increase of $R$ with rotation speed for rough cylindrical walls indicates an increasing effect of
the rough cylindrical wall on 
particle trajectories. This
is likely due to the fact that particle trajectories are almost perpendicular
to the cylindrical wall while they are tangent to endwalls.

The effect of the fill level is more complicated. When increasing the fill level, the endwall cell reduces its size moderately for a smooth wall tumbler and increases its size for a rough wall tumbler (Fig.~\ref{driftvsfill}). For both cases, the
surface topographies are almost unaffected by the fill level. Thus, $D_{ew}$ does not change with fill level. The curvature decreases with increasing fill level near the endwall for both the smooth and rough wall tumblers
(Fig.~\ref{curvefill}). As the fill level increases from 25 to 50\%, the $R$ coefficient increases from 12.8 to 16.2 for smooth walls but decreases from 30.3 to 23.2 for rough walls (see Figs. S16 and S17  \cite{Supplemental}). The variation in $R$ related to curvature for smooth walls dominates, giving an increase of $D_{cyl}$ and an endwall cell size reduction. For rough walls, $R$ and curvature dependence on fill level increases the endwall cell size. 
The variation in $R$ may come about because of the combination of the angle at which particles in the flowing layer impact the downstream wall, which is perpendicular for 50\% fill level and oblique for smaller fill levels, and the roughness of the wall, which tends to hold the particles in place strongly for thin layer as they arrive at the downstream end of the flowing layer. These two effects oppose one another and both vary with  different amplitudes for rough and smooth tumblers.
Friction with particles at rest is probably the dominant effect for rough tumblers and can explain the decrease in $R$ with increasing fill level.
The inclination of trajectory dominates for a smooth tumbler, leading to an increase of $R$ with increasing fill level.

The case of mixed wall roughness confirms the complex dependence of $D_{cyl}$ on fill level (Figs.~S18 to S20 \cite{Supplemental}). Similar to fully smooth and fully rough tumblers, the topography is unaffected by fill level variations, 
 so $D_{ew}$ does not vary with fill level. For a rough cylindrical wall and smooth endwalls, the curvature is almost unaffected by the fill level (Fig.~S19  \cite{Supplemental}). Hence, the slight increase of the endwall cell size with decreasing fill level is only due to an increase of the $R$ coefficient. The same increase of $R$ is expected for the fully rough tumbler, but as the rough endwall induces an increase of the curvature with decreasing fill level, both effects combine to produce a strong decrease in the size of the endwall cell. For a tumbler with a smooth cylindrical wall and rough endwalls, the curvature variation with fill level
is similar to that of a fully smooth tumbler (Fig. S19  \cite{Supplemental}). The variation of the $R$ coefficient with fill level is small for a tumbler with rough endwalls and a smooth cylindrical wall (Fig. S17  \cite{Supplemental}). As a consequence, the decrease of the endwall cell with fill level is small and similar to that of a fully smooth tumbler (Fig.~S18  \cite{Supplemental}).

\section{Conclusions}

 Recirculation cells in cylindrical tumblers are a consequence of the axial displacement and the associated drifts created by the friction on both the tumbler endwalls and the tumbler cylindrical wall. Endwall friction induces trajectory curvature and drift of surface particles toward the endwall, and increasing roughness induces more curved particle
trajectories, while cylindrical wall roughness reduces the curvature (Fig.~\ref{trajmixed}).  On the other hand, axial particle drift is mainly controlled by the cylindrical wall roughness. For the endwall cells, a smooth cylindrical wall enhances drift, while a rough cylindrical wall reduces drift (Fig.~\ref{trajmixed}). For long enough tumblers, the opposite occurs for central cells. In either case, this drift induces recirculation cells, with the endwall cell having a surface flow directed toward the endwall, and a central recirculation cell having surface flow directed toward the center of the tumbler (Figs.~\ref{displacementmap}
and~\ref{maphybbyhsurf}). 

 The concept of two opposing drifts, endwall-induced drift linked to the free surface topography and cylindrical wall-induced drift linked to both the trajectory curvature and a coefficient taking into account the cylindrical wall roughness, provides a framework for understanding the effects of wall roughness, rotation speed, and fill level on the recirculation cells. Endwall
drift dominates near the endwall, inducing the recirculation cell with 
a surface flow directed toward the endwall. Both drifts decrease moving
toward the tumbler center, but the endwall effect decreases more rapidly than the cylindrical effect. For long enough tumblers, drift due to the cylindrical wall
can become dominant, resulting in a new pair of recirculation cells adjacent to the endwall cells. Since endwall cells are due
to the endwall drift dominating and central cells are due
to the cylindrical drift dominating, there is no hydrodynamical coupling between them and no additional cells appear in longer tumblers. In spherical and double-cone tumblers, the trajectory curvatures are reversed compared to the cylindrical tumbler case, so both drifts are directed toward
the pole. As a result, only one pair of cells is observed in these geometries.

These results have implications for both studies of and applications for cylindrical
tumblers.  While it has been known that curved particle trajectories near
the endwalls extend about $D/2$ from the endwalls \cite{ChenOttino08,PohlmanMeier06} and that
recirculation cells related to axial drift can occur that extend even
further from the endwalls \cite{DOrtonaThomas18}, here we confirm those
results in terms of
axial drift and particle trajectory curvature, considering the additional
effect of wall roughness.  The implication is that wall roughness and
tumbler length matter in studies of granular flows in tumblers as well as
their application in industry.  In fact, cylindrical wall roughness is
sometimes intentionally added to industrial tumblers in the form of
``lifters," which are axially oriented bars that protrude slightly from the
inner wall of the cylinder to prevent slip of the bed of particles with
respect to the tumbler.  The results in this study provide useful
guidelines for the study of granular flow in tumblers, both experimental
work and DEM simulations, as well as the practical implications of smooth and rough
walls in applications of tumblers in industry.

\section*{Acknowledgments}
RML and UDO thank CNRS-PICS for financial support. Centre de Calcul Intensif d'Aix-Marseille University is acknowledged for granting access to its high performance computing resources.



\begin{thebibliography}{0}%
\makeatletter
\providecommand \@ifxundefined [1]{%
 \@ifx{#1\undefined}
}%
\providecommand \@ifnum [1]{%
 \ifnum #1\expandafter \@firstoftwo
 \else \expandafter \@secondoftwo
 \fi
}%
\providecommand \@ifx [1]{%
 \ifx #1\expandafter \@firstoftwo
 \else \expandafter \@secondoftwo
 \fi
}%
\providecommand \natexlab [1]{#1}%
\providecommand \enquote  [1]{``#1''}%
\providecommand \bibnamefont  [1]{#1}%
\providecommand \bibfnamefont [1]{#1}%
\providecommand \citenamefont [1]{#1}%
\providecommand \href@noop [0]{\@secondoftwo}%
\providecommand \href [0]{\begingroup \@sanitize@url \@href}%
\providecommand \@href[1]{\@@startlink{#1}\@@href}%
\providecommand \@@href[1]{\endgroup#1\@@endlink}%
\providecommand \@sanitize@url [0]{\catcode `\\12\catcode `\$12\catcode
  `\&12\catcode `\#12\catcode `\^12\catcode `\_12\catcode `\%12\relax}%
\providecommand \@@startlink[1]{}%
\providecommand \@@endlink[0]{}%
\providecommand \url  [0]{\begingroup\@sanitize@url \@url }%
\providecommand \@url [1]{\endgroup\@href {#1}{\urlprefix }}%
\providecommand \urlprefix  [0]{URL }%
\providecommand \Eprint [0]{\href }%
\providecommand \doibase [0]{http://dx.doi.org/}%
\providecommand \selectlanguage [0]{\@gobble}%
\providecommand \bibinfo  [0]{\@secondoftwo}%
\providecommand \bibfield  [0]{\@secondoftwo}%
\providecommand \translation [1]{[#1]}%
\providecommand \BibitemOpen [0]{}%
\providecommand \bibitemStop [0]{}%
\providecommand \bibitemNoStop [0]{.\EOS\space}%
\providecommand \EOS [0]{\spacefactor3000\relax}%
\providecommand \BibitemShut  [1]{\csname bibitem#1\endcsname}%
\let\auto@bib@innerbib\@empty
\end{thebibliography}%


\begin{thebibliography}{99}
\bibitem{GDRMidi04} GDR MiDi, Eur. Phys. J. E {\bf 14}, 341 (2004).
\bibitem{OttinoKhakhar00}J. M. Ottino and D. V. Khakhar, Annu. Rev. Fluid
Mech. {\bf 32}, 55 (2000).
\bibitem{MeierLueptow07}S. W. Meier, R. M. Lueptow, and J. M. Ottino, Adv. Phys. {\bf 56}, 757 (2007).
\bibitem{OrpeKhakhar04}A. V. Orpe and D. V. Khakhar, Phys. Rev. Lett.
{\bf 93}, 068001 (2004).
\bibitem{JainOttino02}N. Jain, J. M. Ottino, and R. M. Lueptow, Phys.
Fluids {\bf 14}, 572 (2002).
\bibitem{OrpeKhakhar07}A. V. Orpe and D. V. Khakhar, J. Fluid Mech. 
{\bf 571}, 1 (2007).
\bibitem{ClementRajchenbach95}E. Clement, J. Rajchenbach, and J. Duran,
Europhys. Lett. {\bf 30}, 7 (1995).
\bibitem{FelixFalk02} G. F\'elix, V. Falk, and U. D'Ortona, Powder Technol. {\bf 128}(2), 314 (2002).
\bibitem{FelixFalk07} G. F\'elix, V. Falk, and U. D'Ortona, Euro. Phys. J. E {\bf 22}, 25 (2007).
\bibitem{PohlmanOttino06} N. A. Pohlman, J. M. Ottino, and R. M. Lueptow, Phys. Rev. E {\bf 74}, 031305 (2006).
\bibitem{SantomasoOlivi04} A. Santomaso, M. Olivi, and P. Canu, Chem. Eng. Sci. {\bf 59}, 3269 (2004).
\bibitem{ChenOttino08}P. Chen, J. M. Ottino, and R. M. Lueptow, Phys. Rev. E {\bf 78}, 021303 (2008).
\bibitem{DOrtonaThomas18} U. D'Ortona, N. Thomas, and R. M. Lueptow, Phys. Rev. E {\bf 97}, 052904 (2018).
\bibitem{BridgwaterSharpe69} J. Bridgwater, N. W. Sharpe, and D. C. Stocker, Trans. Inst. Chem.  Eng. {\bf 47}, T114 (1969).
\bibitem{HillKakalios94}K. M. Hill and J. Kakalios, Phys. Rev. E {\bf 49}, R3610 (1994).
\bibitem{FiedorOttino03}S. J. Fiedor and J. M. Ottino, Phys. Rev. Lett. {\bf 91}, 244301 (2003).
\bibitem{ManevalHill05} J. E. Maneval, K. M. Hill, B. E. Smith, A. Caprihan, and E. Fukushima, Granular Matter {\bf 7}, 199 (2005).
\bibitem{PohlmanMeier06} N. A. Pohlman, S. W. Meier, R. M. Lueptow, and J. M. Ottino, J. Fluid Mech. {\bf 560}, 355 (2006).
\bibitem{ZamanDOrtona13}Z. Zaman, U. D'Ortona, P. B. Umbanhowar, J. M. Ottino, and R. M. Lueptow, Phys. Rev. E {\bf 88}, 012208 (2013).
\bibitem{DOrtonaThomas15}U. D'Ortona, N. Thomas, Z. Zaman, and R. M. Lueptow,
Phys. Rev. E {\bf 92}, 062202 (2015).
\bibitem{ChenLueptow10}P. Chen, J. M. Ottino, and R. M. Lueptow, Phys. Rev. Lett. {\bf 104}, 188002 (2010).
\bibitem{DOrtonaThomas16}U. D'Ortona, N. Thomas, and R. M. Lueptow, Phys. Rev. E {\bf 93}, 022906 (2016).
\bibitem{YuLueptow20}M. Yu, J. M. Ottino, R. M. Lueptow, P. B. Umbanhowar, Phys. Rev. E, submitted, 2021.
\bibitem{KrishnarajNott15} K.P. Krishnaraj and P. R. Nott, Nature Comm. {\bf 7},
10630 (2016).
\bibitem{BroduRichard13} N. Brodu, P. Richard and R. Delannay, Phys. Rev. E {\bf 87}, 022202 (2013).
\bibitem{BroduDelannay15} N. Brodu, R. Delannay, A. Valance and P. Richard  J. Fluid Mech. {\bf 769}, 218 (2015).
\bibitem{ForterrePouliquen01} Y. Forterre and O. Pouliquen, Phys. Rev. Lett. {\bf 86}, 5886 (2001).
\bibitem{ForterrePouliquen02}Y. Forterre and O. Pouliquen,  J. Fluid Mech. {\bf 467}, 361 (2002).
\bibitem{BorzsonyiEcke09} T. B\"orzs\"onyi, R. E. Ecke, and J. N. McElwaine, Phys. Rev. Lett. {\bf 103}, 178302 (2009).
\bibitem{DOrtonaThomas20} U. D'Ortona and N. Thomas, Phys. Rev. Lett.  {\bf 124}, 178001 (2020).
\bibitem{CundallStrack79} P. A. Cundall and O. D. L. Strack, Geotechnique {\bf 29}, 47 (1979).
\bibitem{Ristow00}G. H. Ristow, Pattern Formation in Granular Materials (Springer-Verlag, Berlin, 2000).
\bibitem{SchaferDippel96}J. Sch\"afer, S. Dippel, and D. E. Wolf, J. Phys. I (France) {\bf 6}, 5 (1996).
\bibitem{DonaldRoseman62}M. B. Donald and B. Roseman, Br. Chem. Eng. {\bf 7}, 749 (1962).
\bibitem{DasguptaKhakhar91} S. Das Gupta, D. V. Khakhar, and S. K.
Bhatia, Chem. Eng. Sci. 46, 1513 (1991).
\bibitem{Nakagawa94}M. Nakagawa, Chem. Eng. Sci. {\bf 49}, 2540 (1994).
\bibitem{HillKakalios95}K. M. Hill and J. Kakalios, Phys. Rev. E {\bf 52}, 4393 (1995).
\bibitem{HillCaprihan97}K. M. Hill, A. Caprihan, and J. Kakalios, Phys.
Rev. E 56, 4386 (1997).
\bibitem{Rapaport02}D. C. Rapaport, Phys. Rev. E {\bf 65}, 061306 (2002).
\bibitem{AllenTildesley02}M. P. Allen and D. J. Tildesley, Computer Simulation of Liquids (Oxford University Press, New York, 2002).
\bibitem{DrakeShreve86}T. G. Drake and R. L. Shreve, J. Rheol. {\bf 30}, 981 (1986).
\bibitem{FoersterLouge94}S. F. Foerster, M. Y. Louge, H. Chang, and K. Allia, Phys. Fluids {\bf 6}, 1108 (1994).
\bibitem{ChenLueptow11}P. Chen, J. M. Ottino, and R. M. Lueptow, New J. Phys. {\bf 13}, 055021 (2011).

\bibitem{TaberletNewey06}N. Taberlet, M. Newey, P. Richard, and W. Losert, J. Stat. Mech. P07013 (2006).
\bibitem{Campbell02}C. S. Campbell, J. Fluid Mech. {\bf 465}, 261 (2002).
\bibitem{SilbertGrest07}L. E. Silbert, G. S. Grest, R. Brewster, and A. J. Levine, Phys. Rev. Lett. {\bf 99}, 068002 (2007).
\bibitem{Supplemental} See Supplemental Material at [href will be inserted by editor].
\bibitem{ChenLueptow09}P. Chen, B. J. Lochman, J. M. Ottino, and R. M. Lueptow, Phys. Rev. Lett., {\bf 102}, 148001 (2009).
\bibitem{HuangLiu13} A. N. Huang, L. C. Liu, and H. P. Juo, Powder Technol. {\bf 239}, 98 (2013)
\end{thebibliography}
\end{document}